\documentclass{article}

\usepackage{setspace}
\usepackage[nonatbib,final]{neurips_2022}
\usepackage[numbers, sort&compress]{natbib}
\usepackage[resetlabels]{multibib}
\newcites{SM}{References}

\usepackage[utf8]{inputenc} % allow utf-8 input
\usepackage[T1]{fontenc}    % use 8-bit T1 fonts
\usepackage{hyperref}       % hyperlinks
\usepackage{url}            % simple URL typesetting
\usepackage{booktabs}       % professional-quality tables
\usepackage{amsfonts}       % blackboard math symbols
\usepackage{nicefrac}       % compact symbols for 1/2, etc.
\usepackage{microtype}      % microtypography
\usepackage{xcolor}         % colors

\usepackage{amsmath}
\usepackage{amsthm}
\usepackage{amssymb}
\usepackage{mathtools}
\usepackage{commath}
\usepackage{mathrsfs}
\usepackage{bbm}
\usepackage{bm}
\usepackage{wrapfig}

\usepackage{comment}

\usepackage[]{algorithm2e}

\newcommand{\R}{\mathbb{R}}
\newcommand{\bn}{\mathbf{n}}

\newcommand{\by}{\mathbf{y}}
\newcommand{\byo}{\mathbf{y^o}}
\newcommand{\byoi}{\mathbf{y}^{\mathbf{o},i}}
\newcommand{\byh}{\mathbf{y^h}}

\newcommand{\pf}{p^{\mathrm{fire}}}

\newcommand{\LL}{\mathcal{L}}
\DeclareMathOperator*{\argmax}{arg\,max}

\title{Mesoscopic modeling of hidden spiking neurons}

\author{%
  Shuqi Wang\thanks{Equal contributions.}~,
  Valentin Schmutz$^*\!$, Guillaume Bellec, Wulfram Gerstner\\
  Laboratory of Computational Neuroscience\\
  École polytechnique fédérale de Lausanne (EPFL) \\
  \texttt{first.lastname@epfl.ch} \\
}

\begin{document}

\maketitle

\makeatletter
\let\newtitle\@title
% \makeatother

\begin{abstract}
    %Since current large-scale electrophysiological recordings only provide a partial picture of the probed neural microcircuit, 
    Can we use spiking neural networks (SNN) as generative models of multi-neuronal recordings, while taking into account that most neurons are unobserved? Modeling the unobserved neurons with large pools of hidden spiking neurons leads to severely underconstrained problems that are hard to tackle with maximum likelihood estimation. In this work, we use coarse-graining and mean-field approximations to derive a bottom-up, neuronally-grounded latent variable model (neuLVM), where the activity of the unobserved neurons is reduced to a low-dimensional mesoscopic description. In contrast to previous latent variable models, neuLVM can be explicitly mapped to a recurrent, multi-population SNN, giving it a transparent biological interpretation. We show, on synthetic spike trains, that a few observed neurons are sufficient for neuLVM to perform efficient model inversion of large SNNs, in the sense that it can recover connectivity parameters, infer single-trial latent population activity, reproduce ongoing metastable dynamics, and generalize when subjected to perturbations mimicking optogenetic stimulation.
  %Spiking neuron models are abstractions whose aim is to bridge the gap between neuronal biophysics and computation. In theory, a mechanistic understanding of brain function could be gained from fitting spiking neural networks (SNN) to multi-neuronal recordings. 
  %Can we use spiking neural networks (SNN) as generative models of multi-neuronal recordings, while taking into account that most neurons are unobserved?  we use coarse-graining and mean-field approximations to derive a bottom-up, mechanistic latent variable model (LVM), where the activity of the unobserved neurons is reduced to a low-dimensional mesoscopic description. The derived LVM -- the Biophysics-informed latent mesosocpic model (BILM) -- can be explicitly mapped to a multi-population SNN, giving it a transparent biophysical interpretation. We show, on synthetic spike trains, that BILM enables efficient model inversion of SNNs, in the sense that it can infer connectivity parameters and single-trial latent population activity, reproduce ongoing metastable dynamics, and generalize when subjected to perturbations mimicking photo-stimulation.
\end{abstract}

\section{Introduction}\label{sec:intro}
The progress of large-scale electrophysiological recording techniques \cite{SteKoc18} begs the following question: can we reverse engineer the probed neural microcircuit from the recorded data? If so, should we try to design large spiking neural networks (SNN), representing the whole microcircuit, capable of generating the recorded spike trains? Such networks would constitute fine-grained mechanistic models and would make \textit{in silico} experiments possible. However appealing this endeavor may appear, it faces a major obstacle -- that of unobserved neurons. Indeed, despite the large number of neurons that can be simultaneously recorded, they add up to a tiny fraction of the total number of neurons involved in any given task \cite{GaoGan15}, making the problem largely underdetermined. Training SNNs with large numbers of hidden neurons is challenging because a huge number of possible latent spike patterns result in the same recurrent input to the recorded neurons, making training algorithms nontrivial \cite{PilLat08, BreSen11, RezWie11,BelWan21}.

%in two respects. First, a huge number of possible latent spike patterns result in the same recurrent input to the recorded neurons, which makes training algorithms nontrivial \cite{PilLat08, BreSen11, RezWie11,BelWan21}. Second, the number of parameters of the SNN (e.g., neuronal time constants, firing thresholds, connectivity, synaptic dynamics) is, a priori, very large, which makes the scientific interpretation of any potential solution unclear.

\begin{figure}
\centering
\includegraphics[width=1\linewidth]{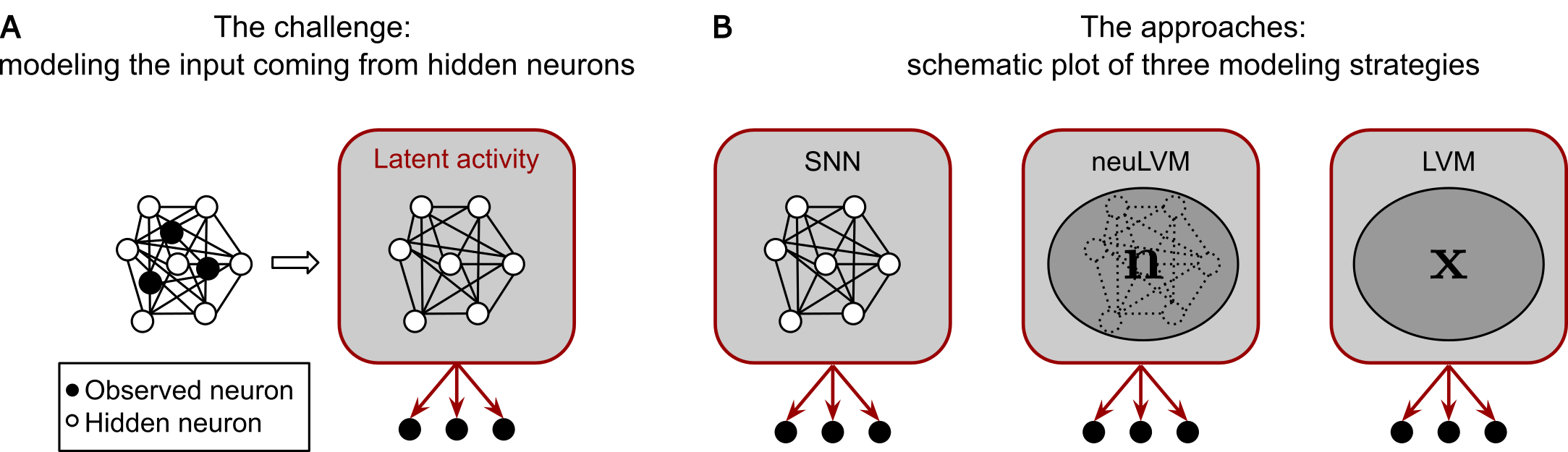}
\caption{\textbf{Training SNNs with large numbers of hidden neurons: challenge and approaches.} 
\textbf{(A)} The challenge of modeling the input to the observed neurons (black) coming from the hidden neurons (white) while only a small fraction of neurons is observed. \textbf{(B)} Modeling strategies for the input coming from the hidden neurons: SNNs (left) model the fine-grained spike trains of all hidden neurons; neuLVM (middle) uses a mesoscopic description of the population activity, clustering neurons into homogeneous populations; classic LVMs (right) model the latent activity with low-dimensional phenomenological variables (the link to SNNs is lost).}\label{fig:intro}
\end{figure}

From the perspective of a single recorded neuron, the spike activity of all the other neurons can be reduced to a single causal variable -- the total recurrent input (Figure~\ref{fig:intro}A). Hence, we argue that fine-grained SNNs are not necessary to model the inputs from hidden neurons but can be replaced by a coarse-grained model of the sea of unobserved neurons. One possible coarse-graining scheme consists in clustering neurons into homogeneous populations with uniform intra- and inter-population connectivity. With the help of mean-field neuronal population equations \cite{Ger95,BruHak99,Ger00,GerKis14}, this approach enables the reduction of large SNNs to low-dimensional mesoscopic models composed of neuronal populations interacting with each other \cite{SchDeg17, SchChi19,SchLoe21}. Clusters can reflect the presence of different cell-types \cite{LefTom09,PotDie14,SchDeg17} or groups of highly interconnected excitatory neurons \cite{YosDan05, SonSjo05, CloBue10, PerBer11, KoHof11, LitDoi12}. From a computational point of view, coarse-grained SNNs offer biologically plausible implementations of rate coding by ensembles of neurons \cite{ShaNew94,ShaNew98} and `computation through neural population dynamics' \cite{VyaGol20}. 

In this paper, we show that, after clustering the unobserved neurons into several homogeneous populations, the finite-size neuronal population equation of \citet{SchDeg17} can be used to derive a neuronally-grounded latent variable model (neuLVM) where the activity of the unobserved neurons is summarized in a coarse-grained mesoscopic description. The hallmark of neuLVM is its direct correspondence to a multi-population SNN. As a result, both model parameters and latent variables have a transparent biological interpretation: the model is parametrized by single-neuron properties and synaptic connectivity; the latent variables are the summed spiking activities of the neuronal populations. Coarse-graining by clustering, therefore, turns an underdetermined problem -- fitting an SNN with a large number of hidden neurons -- into a tractable problem with interpretable solutions.

Switching metastable activity patterns that are not stimulus-locked have attracted a large amount of attention in systems neuroscience \cite{Mil16, LacFon19, BriYan22} for their putative role in decision-making \cite{LatYat15}, attention \cite{EngSte16}, and sensory processing \cite{MazLac19}. Since generative SNN models of these metastable dynamics are available \cite{MorRin07, LitDoi12, MazFon15, SchDeg17}, metastable networks constitute ready-to-use testbeds for bottom-up mechanistic latent variable models.  Therefore, we propose metastable networks as a new benchmark for mechanistic latent variable models.
%The method is tested on synthetic data generated by multi-population SNNs exhibiting metastable dynamics. In these settings, sampling-based strategies for fitting SNNs with hidden neurons like that of \citet{BelWan21} are inapplicable because the latent population activity is not stimulus-locked. Metastable networks constitute ideal testbeds for bottom-up, mechanistic LVM's since fine-grained generative SNN models are available \cite{MorRin07, LitDoi12, MazFon15, SchDeg17}, and .

\section{Relation to prior work}

While many latent variable models (LVM), including Poisson Linear Dynamical Systems (PLDS) \cite{MacBue12} and Switching Linear Dynamical Systems (SLDS) \cite{AckFu70,GhaHin00,Bar06,FoxSud08,PetYu11,LinJoh17}, have been designed for inferring low-dimensional population dynamics \cite{KulPan07, LawWu10,VidAhm12,GaoArc16, WuRoy17,PanOsh18,DunSah18,DunBoh19,NasLin19,GlaWhi20,ZolPil20,RutBer20,KeeAoi20,KimLuo21}, their account of the population activity is a phenomenological one. By contrast, the LVM derived in this work is a true multiscale model, as latent population dynamics directly stems from neuronal dynamics. 
% SLDS: \cite{AckFu70,GhaHin00,Bar06,FoxSud08,PetYu11,LinJoh17}
% inferring low-dimensional population dynamics \cite{KulPan07, LawWu10,VidAhm12,GaoArc16, WuRoy17,PanOsh18,DunSah18,DunBoh19,NasLin19,GlaWhi20,ZolPil20,RutBer20,KeeAoi20,KimLuo21}

Our method builds on \citet{RenLon20}, who showed that the mesoscopic model of \citet{SchDeg17} enables the inference of neuronal and connectivity parameters of multi-population SNNs via likelihood-based methods when the mesoscopic population activity is \textit{fully observable}. Here, for the first time, we show that mesoscopic modeling can also be applied to unobserved neurons, relating LVMs to mean-field theories for populations of spiking neurons \cite{Ger95,BruHak99,Ger00,GerKis14,SchDeg17,SchChi19}. Our neuLVM approach towards unobserved neurons differs from the Generalized Linear Model (GLM) approach \cite{Pan04, TruEde05, PilShl08} (and recent extensions \cite{ZolPil18,LamTul18, KobKur19}), which either neglects unobserved neurons or replaces unobserved neurons by stimulus-locked inputs. Our approach also avoids microscopic simulations of hidden spiking neurons \cite{PilLat08, BreSen11,BelWan21}, which scale poorly with the number of hidden neurons.

\section{Background: mesoscopic modeling of the population activity}\label{sec:meso-theory}

Biophysical neuron models can be accurately approximated (neglecting the nonlinearity of dendritic integration) by simple spiking \textit{point} neurons \cite{KisGer97, JolLew04, BreGer05}, which can be efficiently fitted to neural data \cite{JolRau06, KobTsu09, PozMen15, TeeIye18}. Stochastic spiking neuron models where the neuron's memory of its own spike history is restricted to the time elapsed since its last spike (its \textit{age}) are said to be of renewal-type.\footnote{Traditional renewal theory in the mathematical literature \cite{Cox62} is restricted to stationary input whereas we use `renewal-type' in the broader sense that also includes time-dependent input.} Examples of renewal-type neurons include noisy leaky integrate-and-fire and `Spike-Response Model 0' neurons \cite{Ger00,GerKis14}.  The dynamics of a homogeneous population of interacting renewal-type neurons can be described, in the mean-field limit, by an exact integral equation \cite{WilCow72,Ger95,Ger00,GerKis14,SchChi19} (see \cite{DemGal15,FouLoe16,Che17} for rigorous proofs). In the case of homogeneous but finite populations, \citet{SchDeg17} derived a stochastic integral equation that provides a mesoscopic description (i.e. a description including finite-size fluctuations) of the population dynamics.

For clarity of exposition, in this section and the next, we focus on the case of a single homogeneous population with no external input. All the arguments presented below can be readily generalized to the case of multiple interacting populations with external input (Appendices~\ref{app:meso} \ref{app:neuLVM} \ref{app:neuLVM_fit_appendix}).

Let us consider a homogeneous SNN of $N$ recurrently connected renewal-type spiking neurons. For $T$ discrete time steps of length $\Delta t$, let $\by \in \{0,1 \}^{N \times T}$ (a $N \times T$ binary matrix) denote the spike trains generated by the $N$ neurons. The fact that the neurons are of renewal-type implies, by definition, that the probability for neuron $i$ to emit a spike at time $t$ can be written $p({y}^i_{t}=1| {\by}_{1:t-1}, \Theta) = \rho^{\Delta t}_{\theta^i}(a^i, \sum_j J^{ij} {\by}^j_{1:t-1})$ where $a^i$ is the age of neuron $i$ (i.e. the number of time steps elapsed since the last spike of neuron $i$), the $J^{ij}$ are the recurrent synaptic weights of the network, and $\theta^i$ are the parameters of neuron $i$. The sum $\sum_j J^{ij} {\by}^j_{1:t-1}$ represents the past input received by neuron $i$ in all time steps up to $t-1$. The superscript $\Delta t$ of the function $\rho^{\Delta t}_{\theta^i}$ indicates that we consider here the discrete-time `escape rate' of the neuron but the transition to continuous time is possible \cite{GerKis14}. The explicit expression for $\rho^{\Delta t}_{\theta^i}$ in the case of leaky integrate-and-fire neurons with `escape noise' (LIF) is given in Appendix~\ref{app:meso}.

A crucial notion in this work is that of a `homogeneous population'. The SNN described above forms a homogeneous population if all the recurrent synaptic weights are identical, that is, $J^{ij}=J/N$ (mean-field approximation), and if all the neurons share the same parameters, that is, $\theta^i=\theta$. In a homogeneous population, all the neurons share the same past input $J\bn_{1:t-1}/N$, where $\bn_{1:t-1} = (n_1, n_2, \dots, n_{t-1})$ denotes the total number of spikes in the population in time steps $1,2,\dots,t-1$ with $n_{t'} = \sum_{i=1}^N y^i_{t'}$ being the total number of spikes in the population at time $t'$. Then, for \textit{any} neuron in the population, the probability to emit a spike at time $t$, given its age $a$, is
\begin{equation}
    p_{t,a}^{\mathrm{fire}} = \rho^{\Delta t}_{\theta}(a, J \bn_{1:t-1}/N).
    \label{eq:neuron-model}
\end{equation}

%Finally, if we define a homogeneous population as a recurrently connected population of neurons with shared weight $J^{ij}=J$ and shared neuron parameters $\theta^i = \theta$, dynamics of a homogenous population of spiking neurons is greatly simplified. In fact the following formula enables a mesoscopic description where the entire network is determined by $\bn_t$ the number of spikes emitted inside the population at time $t$.
%Formally, the firing probability given the age $a^i$ and ${\bn}_t$ becomes independent of the neuron index $i$ and we can define the neuron firing probability $p_{t,a}^{\mathrm{fire}}$:
Importantly, Eq.~\eqref{eq:neuron-model} is independent of the identity of the neuron. 

In a microscopic description of the spiking activity, the vector $\by_t$ depends nonlinearly on the past $\by_{1:t-1}$. A mesoscopic description aims to reduce the high-dimensional microscopic dynamics to a lower-dimensional dynamical system involving the population activity $n_t$ only (in the case of multiple interacting populations, $\bn_t$ is a vector of dimension $K$ equal to the number of populations, Appendix~\ref{app:meso}). While an exact reduction is not possible in general (neuron models being nonlinear), a close approximation in the form of a stochastic integral equation was proposed by \citet{SchDeg17}. In discrete time, the stochastic integral equation reads

\begin{subequations}\label{eq:stochastic_integral}
\begin{align}
    % \bn_t &\sim \text{Binomial}(N, \bar{A}_t \Delta t),\label{eq:binomial}\\
    % \bar{A}_t &= \frac{1}{N}\Bigg[\sum_{a\geq 1}\lambda_{t,a}S_{t,a}\,\bn_{t-a} + \underbrace{\Lambda_t\bigg(N - \sum_{a \geq 1}S_{t,a}\,\bn_{t-a}\bigg)}_{\text{`finite-size correction'}}\Bigg]_+.
    n_t &\sim \text{Binomial}\left(N,\; \bar{n}_t /N\right),\label{eq:binomial}\\
    \bar{n}_t &= \Bigg[\sum_{a\geq 1}p^{\mathrm{fire}}_{t,a}S_{t,a}\,n_{t-a} + \underbrace{\Lambda_t\bigg(N - \sum_{a \geq 1}S_{t,a}\,n_{t-a}\bigg)}_{\text{`finite-size correction'}}\Bigg]_+.
\end{align}
\end{subequations}

% The variable $\bar{A}_t$ can be interpreted as the population average firing rate. The conditional intensity $\lambda_{t,a}$ defines the probability for a neuron to spike at time $t$, given that its last spike was emitted $a$ steps ago.
The variable $\bar{n}_t$ can be interpreted as the expected number of neurons firing at time $t$. 
%The conditional firing probability $p^{\mathrm{fire}}_{t,a}$ defines the probability for a neuron to spike at time $t$, given that its last spike was emitted $a$ steps ago.
The survival $S_{t,a}=\prod_{s=0}^{a-1}(1 - \pf_{t-a+s,s})$ is the probability for a neuron to stay silent between time $t-a$ and $t-1$. 
The finite-size correction term stabilizes the model by enforcing the approximate neuronal mass conservation $\sum_{a \geq 1}S_{t,a}\,n_{t-a} \approx N$ (see \cite{SchLoe21} for an in-depth mathematical discussion). The `modulating factor' $\Lambda_t$ has an explicit expression \cite{SchDeg17,SchLoe21} in terms of 
% $\lambda$, 
$p^{\mathrm{fire}}$, $S$ and $\bn$ (indices are dropped here for simplicity, complete formulas are presented in Appendix~\ref{app:meso}, as well as explanations on how to initialize Eq.~\eqref{eq:stochastic_integral}). Importantly, for a populations of \textit{interacting} neurons, 
% $\lambda$, 
$p^{\mathrm{fire}}$, $S$ and $\Lambda$ depend on $\bn$, which makes the stochastic equation \eqref{eq:stochastic_integral} highly nonlinear. While the mesoscopic model~\eqref{eq:stochastic_integral} is not mathematically exact, it provides an excellent approximation of the first and second-order statistics of the population activity \cite{SchDeg17}, and is much more tractable than the exact `field' equation \cite{Che17b,DumPay17}. Also, the mesoscopic model~\eqref{eq:stochastic_integral} can be generalized to the case of non-renewal neurons with spike-frequency adaptation \cite{SchDeg17} and short-term synaptic plasticity \cite{SchGer20}.

Formally, Eq.~\eqref{eq:stochastic_integral} is reminiscent of the Point Process Generalized Linear Model (GLM) \cite{Pan04, TruEde05, PilShl08} for single neurons, with the notable difference that Eq.~\eqref{eq:stochastic_integral} contains additional nonlinearities beyond those of the GLM because $p^{\mathrm{fire}}$, $S$ and $\Lambda$ all depend on $\bn$ (Appendix~\ref{app:meso}). Importantly, Equation~\eqref{eq:stochastic_integral} readily defines an expression for the probability $p(\bn|\Theta)$ \cite{RenLon20}, where $\Theta=\{J,\theta\}$ denotes the model parameters. Thus, the mesoscopic model~\eqref{eq:stochastic_integral} allows us to avoid the intractable sum encountered if we naively try to derive $p(\bn|\Theta)$ directly from the microscopic description (the intractable sum stems from the fact that the identity of neurons is lost in the observation $n_{t'}$ at each time step $t'$, Figure~\ref{fig:intro}B). 
%The use of the mesoscopic model \eqref{eq:stochastic_integral} for inference was pioneered by \citet{RenLon20}, who showed, on synthetic data, that likelihood-based methods can be used to estimate single-neuron and network connectivity parameters from the \textit{observed} population activity.
%in networks of up to four interacting populations. 
%Their method scales up to a model with four interacting populations, adapted from the Potjans-Diesman microcircuit \cite{PotDie14}. 

\section{Theoretical result: Neuronally-grounded latent variable model}\label{sec:neuLVM}
%How can the theory of the previous section help us to identify neuronal and connectivity parameters if only a small fraction of neurons is observed? 
In this section, we first recall why training SNN with large numbers of hidden neurons via the maximum likelihood estimator is computationally expensive. Then, we show that the mesoscopic description, Eq.~\eqref{eq:stochastic_integral}, allows us to derive a tractable, neuronally-grounded latent variable model, which can be mapped to a multi-population SNN.
%Given some spike train recording we attempt now to reconstruct the underlying biological network by fitting a network model to the data. 

For the sake of simplicity, all the arguments are presented for a single homogeneous population, but the generalization to multiple interacting populations is straightforward (Appendices~\ref{app:neuLVM} and \ref{app:neuLVM_fit_appendix}). Let us assume that we observe, during $T$ time steps, the spike trains of $q$ simultaneously recorded neurons that are part of a homogeneous population of $N$ neurons, with $N>q$. We split the spike trains of the entire population $\by\in\{0,1\}^{N\times T}$ into the observed spike trains $\byo$ ($q$ neurons) and hidden spike trains $\byh$ ($N-q$ neurons). Even for a single population, it is difficult to infer the parameters $\Theta = \{J,\theta\}$ of the SNN from observation $\byo$ using the maximum likelihood estimator because, in the presence of hidden neurons, the likelihood $\LL$ involves a marginalization over the latent spike trains $\byh$:
\begin{equation} \label{eq:L_for_SNN}
    \LL = p(\byo | \Theta)  = \sum_\byh p(\byo, \byh | \Theta).
\end{equation}

While different variants of the Expectation-Maximization (EM) algorithm \cite{DemLai77} relying on sampling $\byh$ have been used to maximize the likelihood \cite{BelWan21, PilLat08, BreSen11}, these algorithms scale poorly with the number of hidden neurons. 

Instead, we exploit the fact that, for a homogeneous population, the fine-grained knowledge of the latent activity $\byh$ is not necessary since all the observed neurons receive at time $t$ the same input $J n_t$, where $ n_t=\sum_{i=1}^N y^i_t$ is the population activity of Section~\ref{sec:meso-theory}. Hence, we rewrite the likelihood~\eqref{eq:L_for_SNN} as

%The maximization problem with latent variables $\byh$ could in principle be addressed with the Expectation-Maximization (EM) algorithm \cite{DemLai77}. However, here, the posterior $p(\byh|\byo, \Theta)$ is non-trivial to estimate when the number of hidden neurons is large. Previous attempts have tried to estimate the posterior $p(\byh|\byo, \Theta)$ via sampling \cite{BelWan21, PilLat08, BreSen11, RezWie11} but these methods scale poorly with the number of hidden neurons.

%Assuming that there some redundancy between the observed and hidden neurons, we propose to use the mesoscopic description of section~\ref{sec:meso-theory} to solve the problem of unobserved neurons.
%Our method, which we name as Biophysics-informed latent mesoscopic model (BILM), relies on assuming that the observed data is sampled from homogeneous populations. Under this assumption, we can coarse-grain neurons in SNN (including $q$ observed and $N-q$ hidden ones) into multiple homogenous populations (Figure~\ref{fig:intro}~B) and rewrite the likelihood $\LL$ (Eq.~\ref{eq:L_for_SNN}) as 
\begin{subequations}\label{eq:L_for_neuLVM}
\begin{equation}
      \LL = p(\byo | \Theta)  = \sum_\bn p(\byo,\bn| \Theta),   
\end{equation}
%where $\bn_t=\sum_{i=1}^N\by_t^{i}$ being a $1\times T$ vector denotes the population activity. 
where the probability $p(\byo,\bn| \Theta)$ factorizes in $T$ terms of the form
\begin{equation}\label{eq:Pyn_facotirzed}
      p(\by^{\mathbf{o}}_t, n_t | \by^{\mathbf{o}}_{1:t-1}, \bn_{1:t-1}, \Theta)  = \underbrace{p(\by^{\mathbf{o}}_t|\by^{\mathbf{o}}_{1:t-1}, \bn_{1:t-1}, \Theta)}_{\text{given by neuron model, Eq.~\eqref{eq:neuron-model}}}\underbrace{p(n_t|\bn_{1:t-1}, \Theta)}_{\text{approx. by meso. model, Eq.~\eqref{eq:stochastic_integral}}}.   
\end{equation}
\end{subequations}

A comparison of Eqs.~\eqref{eq:L_for_SNN} and \eqref{eq:L_for_neuLVM} shows that the high-dimensional latent activity $\byh$ has been reduced to a low-dimensional mesoscopic description. Importantly, the $q$ observed spike trains are conditionally independent given the population activity $\bn$. While the conditional dependence structure implied by Eq.~\eqref{eq:Pyn_facotirzed} is typical of standard latent variable models of multi-neuronal recordings \cite{SmiBro03,KulPan07,YuCun09,MacBue12}, in our approach, the latent variable explicitly represents the population activity of the generative SNN and the parameters of the model are identical to those of the SNN. As the latent population dynamics directly stems from neuronal dynamics, we call our LVM the neuronally-grounded latent variable model (neuLVM).
%The probability $p(\bn_t|\bn_{1:t-1}, \Theta)$ can only be approximated by the mesoscopic model because, as mentioned in Section~\ref{sec:meso-theory}, the mesosocpic model is not exact. However, extensive simulations have shown that the model~\eqref{eq:stochastic_integral} gives an accurate approximation of the true population dynamics and it is exact in the limit $N\to\infty$ \cite{SchDeg17}. 

The nonlinearity and the non-Markovianity of Eq.~\eqref{eq:stochastic_integral} prevent us from using previous EM algorithms \cite{SmiBro03,KulPan07,YuCun09, MacBue12}. Therefore, we fit the neuLVM via the Baum-Viterbi algorithm \cite{EphMer02} (also known as Viterbi training or hard EM \cite{AllGal11}), which alternates estimation (E) and maximization (M) step

%Given the observed spike trains $\byo$, fitting BILM can be done via a EM-like algorithm. Since the mesoscopic dynamics \eqref{eq:stochastic_integral} is nonlinear and non-Markovian, EM-type algorithms like that in \cite{SmiBro03,KulPan07,YuCun09} cannot be applied. Instead, we use here a variational approach consisting in approximating the posterior $p(\bn|\byo,\Theta)$ by a point mass $\delta_\mu$, where $\mu=\argmax_{\bn}\log p(\byo, \bn|\Theta)$. This approximation relies on the heuristic that the posterior $p(\bn|\byo,\Theta)$ is concentrated around its maximum. For the $n$-th iteration, the \textbf{E-step} and the \textbf{M-step} read

\begin{enumerate}
    \item[\textbf{E-step}.] $\widehat{\bn}^{n} = \operatorname{argmax}_\bn \log p(\byo,\bn| \widehat{\Theta}^{n-1})$,
    % \item[\textbf{Step E}.] Optimize $\bn$ in $\log p(\byo,\bn| \widehat{\Theta}^{n})$ with gradient ascent. Write $\widehat{\bn}^{n+1}$ the result of this optimization step
    \item[\textbf{M-step}.] $\widehat{\Theta}^{n} = \operatorname{argmax}_\Theta \log p(\byo,\widehat{\bn}^{n}| \Theta)$.
    % \item[\textbf{Step M}.] Optimize $\Theta$ in 
    %     $\log p(\byo,\widehat{\bn}^{n+1}| \Theta)$ 
    %     with gradient ascent. Write $\widehat{\Theta}^{n+1}$ the result of this optimization step.
\end{enumerate}
The estimated parameters $\widehat{\Theta}$ and the estimated latent population activity $\widehat{\bn}$ are the result of many iterations of \textbf{E-step} and \textbf{M-step} (Appendix~\ref{app:neuLVM_fit_appendix}). Note that the computational cost of this algorithm does not depend on the number of hidden neurons (it only depends on the number of populations).\footnote{Our implementation of the algorithm is openly available at {\tt \url{https://github.com/EPFL-LCN/neuLVM}}.}.

\section{Experimental results}

% \subsection{Mean-field models can generate rich network dynamics}
\subsection{Single homogenous population: SNN with metastable cluster states}\label{sec:single}

Although seemingly simple, homogeneous populations of leaky integrate-and-fire (LIF) neurons without external stimulation are SNNs with a rich repertoire of population dynamics, including asynchronous states, synchronous states, and cluster states \cite{Ger95,Ger00}. 
%Hence, they are the first testbed for our neuLVM approach.
%A homogeneous population of leaky integrate-and-fire (LIF) neurons without external stimulation can already produce a rich repertoire of activity patterns, e.g., asynchronous states, synchronous states, and cluster states \cite{Ger95,Ger00}. Moreover, finite-size fluctuations can make these states metastable, induce switches between metastable states, and break the symmetry of unstable states. All these complex behaviors are faithfully reproduced by the mesoscopic model~\eqref{eq:stochastic_integral}. 
%To illustrate the expressivity of these seemingly simple generative models and the power of the inductive biases contained in neuLVM, we focus our experiments on cluster states. 
In a $m$-cluster state (with $m\geq 2$), the population activity oscillates at a frequency $m$ times higher than the average neuronal firing rate: a neuron spikes every $m$ cycle on average; conversely, approximately $N/m$ neurons fire in each cycle ($N$ being to the total number of neurons). Cluster states have therefore been described as `higher harmonics' of the synchronous state (or $1$-cluster state) \cite{GerHem93, GolRin94, ErnPaw95, Ger00} where all neurons fire in synchrony. 

%The neuLVM relies on homogeneous populations to abstract away the neurons which are not recorded, but even a single homogeneous population of leaky integrate-and-fire (LIF) neurons without external stimulation can produce rich patterns of population activity, such as cluster states. 

\begin{figure}
\centering
\includegraphics[width=1\linewidth]{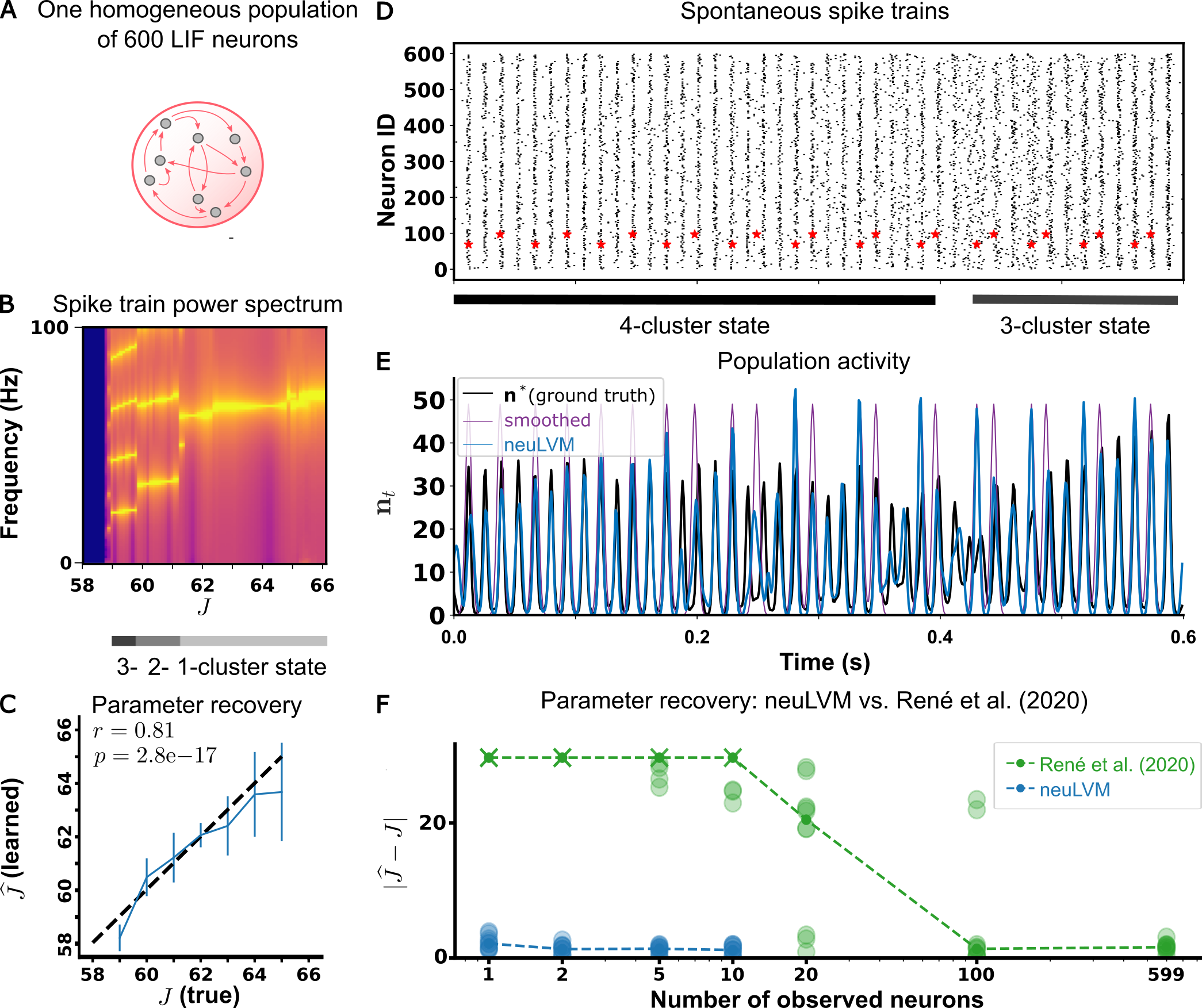}
\caption{\textbf{Single-population SNN with metastable cluster states.} \textbf{(A)} Network architecture (for visualization purposes, only a few connections are drawn).
\textbf{(B)} Spike train power spectrum for different choices of connectivity parameter $J$. All simulations start from the same unstable asynchronous state. The corresponding cluster states are indicated below. The blue region around $J=58$ mV indicates the absence of activity.
\textbf{(C)} Connectivity recovered by the neuLVM $\widehat{J}$ vs ground truth $J$. The neuLVM was fitted on one-second single-trial recordings of six neurons ($1\%$ of the population). For each ground truth $J$ value (seven in total), ten different trials were generated: bars indicate the standard deviations of the recovered $\widehat{J}$. The Pearson correlation coefficient between the recovered $\widehat{J}$ and $J$ is $r=0.81$ and the associated p-value is $2.8\mathrm{e}{-17}$ (see Table~\ref{tab:1pop-result-fit}).
\textbf{(D)} Spike trains generated by the ground truth SNN for a trial showing a transition from a metastable $4$-cluster state to a $3$-cluster state. The spike trains of two randomly sampled neurons (red stars) formed the training data (for visualization purposes, only the first $0.6$ second of the one-second trial is shown) on which neuLVM was fitted: \textbf{(E)} the inferred population activity $\widehat{\bn}|\byo$ is compared to the ground truth $\bn^*$ and the summed, smoothed spike trains (Gaussian smoothing window with $\sigma = 1.4$ ms, Appendix~\ref{app:smoothing}) of the two observed neurons.
\textbf{(F)} Absolute difference between the recovered $\widehat{J}$ and the ground truth $J$ for the neuLVM algorithm and the method of René et al. (2020) for varying numbers of observed neurons. Using the same trial as in D, for each number of observed neurons, the two methods were tested on $10$ different samples of observed neurons (see Table~\ref{tab:1pop-result-comp}). The marker `$\times$' indicates that the difference $|\widehat{J}-J|$ is larger than $30$ mV. The median samples are linked with dashed lines to show the trends.
%The connectivity $\widehat{J}$ inferred by the neuLVM and \citet{RenLon20}, as the number of observed neurons varies. `x' means the error $|\widehat{J}-J|$ is larger than 30. The medians of errors are denoted with the dots and linked to show the trends. For each number of observed neurons, ten different datasets were generated and tested. One of the blue dots was trained to the recorded spike trains as in D.
}\label{fig:1pop-exp}
\end{figure}

In this set of experiments, we always consider the same network of $600$ LIF neurons (Figure~\ref{fig:1pop-exp}A), where only the connectivity parameter $J$ varies. When initialized at time $0$ in the same unstable asynchronous state, the network can spontaneously settle in a $m$-cluster state, where $m$ depends on the recurrent connectivity parameter $J$  (Figure~\ref{fig:1pop-exp}B): finite-size fluctuations break the symmetry of the asynchronous state and the population splits into $m$ groups of synchronized neurons.
%This phenomenon, which is fully captured by the mesoscopic model~\eqref{eq:stochastic_integral}, is due to the fluctuations-induced symmetry breaking of the asynchronous state and the conse splitting of the population into $m$ groups of synchronized neurons. 
The cluster state to which the network converges can be read from the power spectrum of the neuronal spike trains (Figure~\ref{fig:1pop-exp}B) (the fundamental frequency of the $m$-cluster state is approximately $m$ times lower than that of the $1$-cluster state). Generating spike trains for $6$ observed neurons ($1\%$ of the population), we tested whether neuLVM could recover the connectivity parameter $J$ (neuronal parameters $\theta$ were given), for different $J$'s in the $1$-, $2$-, and $3$-cluster states range (Figure~\ref{fig:1pop-exp}C, Table~\ref{tab:1pop-result-fit}). 
%To make sure that our method was learning $J$, the Baum-Viterbi algorithm was initialized with $\widehat{J}^0$ sampled outside of the interval $[30,90]$. 
The Pearson correlation between the learned $\widehat{J}$ and the true $J$ was $0.81$ with p-value $2.8\mathrm{e}{-17}$, showing that, statistically, neuLVM could recover the connectivity parameter of the SNN.
%This complex behavior can be fully captured by the mesoscopic model~\eqref{eq:stochastic_integral}, highlighting the fact that, in contrast to standard rate models \cite{WilCow72,JanRit95,DecJir08}, Eq.~\eqref{eq:stochastic_integral} is high-dimensional stochastic dynamical system \cite{SchDeg17,SchLoe21}.

To assess how well neuLVM can infer the latent population activity and how neuLVM compares with the methods assuming full observability (like \citet{RenLon20}), we studied in detail a single trial showing a transition from a metastable $4$-cluster state to a $3$-cluster state (Figure~\ref{fig:1pop-exp}D,E). To generate this trial, we chose $J = 60.32$ mV and initialized the network in the $4$-cluster state. From the spike trains of only two neurons (red stars in Figure~\ref{fig:1pop-exp}D), neuLVM could infer the ground truth population activity $\bn^*$ during the $4$-cluster state,  and during the $3$-cluster state, and could approximately detect the transition between the two states (Figure~\ref{fig:1pop-exp}E). 
%The neuronal parameters $\theta$ were given and the Baum-Viterbi algorithm was initialized with $\widehat{\bn}^0$ being the summed smoothed spike trains of the two observed neurons (purple curve in Figure~\ref{fig:1pop-exp}~E). 
While the summed, smoothed spike trains missed two out of four population activity peaks in the $4$-cluster state, and one out of three peaks in the $3$-cluster state (purple curve in Figure~\ref{fig:1pop-exp}E), the strong inductive biases contained in neuLVM enabled the inference of the `missing' peaks (blue curve in Figure~\ref{fig:1pop-exp}E). Finally, neuLVM and a method assuming full observability (equivalent to a naive application of \citet{RenLon20}) were compared through their ability to recover the connectivity parameter $J$, for varying numbers of observed neurons (Figure~\ref{fig:1pop-exp}F, Table~\ref{tab:1pop-result-comp}). Since a naive application of the method of \citet{RenLon20} does not take into account hidden neurons, it led, as expected, to wildly inaccurate estimates $\widehat{J}$ when the summed spike train was far from the ground truth population activity (which happened when the number of observed neurons was small, see Figure~\ref{fig:1pop-exp}E for an example). In contrast, the neuLVM managed to recover the connectivity parameter, thanks to the fact that the Baum-Viterbi algorithm of Section~\ref{sec:neuLVM} also infers the population activity (see Figure~\ref{fig:1pop-exp}E for an example). 

\subsection{Multiple populations: SNN with metastable point attractors}\label{sec:3pop}

\subsubsection{Latent population activity inference and reproduction of metastable dynamics} \label{sec:reproduction}

\begin{wrapfigure}{r}{0.50\textwidth}
  \begin{center}
    \raisebox{0pt}[\dimexpr\height-0.6\baselineskip\relax]{\includegraphics[width=0.45\textwidth]{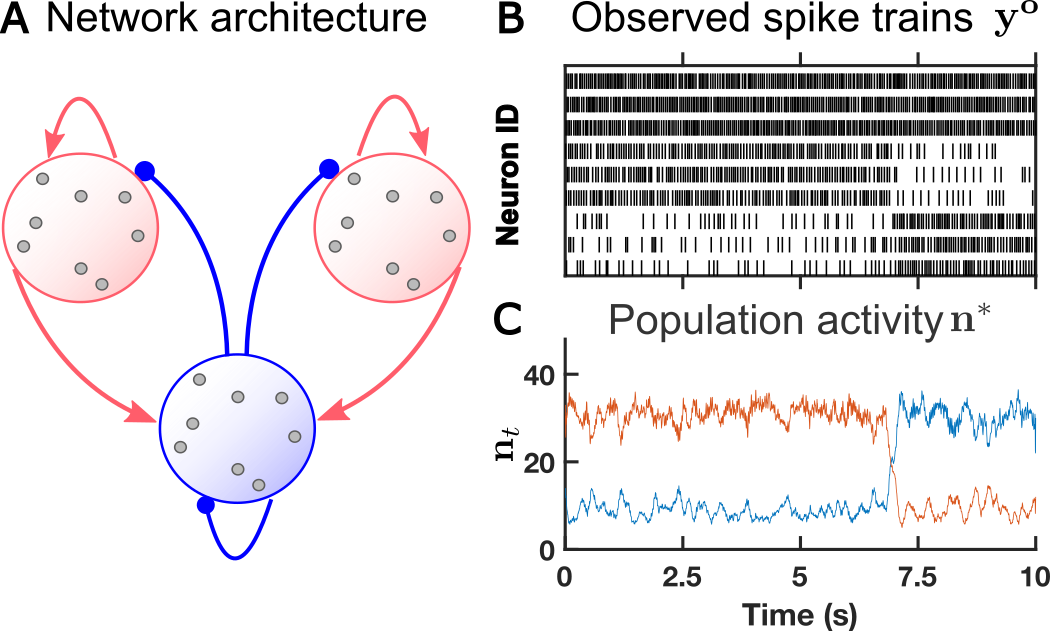}}
  \end{center}
  \caption{\textbf{Network architecture and example trial.} \textbf{(A)} Architecture of the three-population, metastable, winner-take-all SNN. \textbf{(B)} Example trial from the spiking benchmark dataset: 10 seconds recordings of $9$ observed neurons ($3$ neurons from each of the three populations) and \textbf{(C)} corresponding ground truth latent population activity (for the two excitatory populations).} \label{fig:3popdata}
\end{wrapfigure}

As a second benchmark, we tested neuLVM on synthetic data generated by three interacting populations with two populations of $400$ excitatory LIF neurons and one population of $200$ inhibitory neurons (Figure~\ref{fig:3popdata}A). Recurrent connections between these population drive winner-take-all dynamics with finite-size fluctuations-induced switches of activity between the two excitatory populations \cite{SchDeg17,WonWan06} -- an example of `itinerancy between attractor states' \cite{Mil16}. The population activities of this metastable, three-population SNN constitute the ground truth against which different models will be tested.

To build a spiking benchmark dataset, 
%we proceed in two steps. In a preprocessing step, we cluster the data to assign the observed neurons to homogeneous populations. To verify the feasibility of
%clustering, we take $M=20$ samples of $q=25$ neurons each, selected
%randomly (with replacement) from the $600$ neurons. We low-pass
%filter the spike trains ($\tau_{\mathrm{cluster}}= 400\,\text{ms}$) and use the
%van-Rossum-Distance \cite{vanRos01} to evaluate pairwise distance. Searching for three
%clusters with k-means clustering \cite{Llo57}, we find that in
%all $M=20$ experiments, the three different clusters correspond to the
%three different populations so that all members of a given cluster
%represented neurons of the same population. Moreover, we found that no
%cluster contained less than three neurons. In a second step, 
we randomly selected $9$ neurons -- $3$ neurons from each of the three populations -- and considered the spike trains of these neurons as the observed data. For simplicity, the correct partitioning of the $9$ neurons into $3$ groups is given since it can be reliably obtained by k-means clustering \cite{Llo57} using the van-Rossum-Distance \cite{vanRos01} between spike trains. 
The complete dataset consists of $20$ trials of $10$ seconds. An example trial is shown in Figure~\ref{fig:3popdata}B.

%The neuLVM is initialized with random parameters in a range [...]. 
%To assess whether BILM can reproduce these dynamics we recorded 3 neuron from each population for a single trial of $T=10$, $40$, or $80$ seconds where at least one transition occurred.%\textcolor{cyan}{($T=10, 40, or 80; as an example, chosen randomly, we show $T$=10 seconds in Figure~\ref{fig:main-exp}~A)}. We say that BILM reproduces the metastable dynamics if, after learning, spike trains generated by free simulations of BILM exhibit the same metastable behavior as the training data. 
%At initialization, the network did not show any metastable behavior (Figure~\ref{fig:ap-init}). After fitting the parameters to the recorded data, spike trains generated by the fitted model exhibit the metastable winner-take-all dynamics with stochastic transitions (an example is shown in Figure~\ref{fig:main-exp}~B C).

\begin{figure}
\centering
\includegraphics[width=0.75\linewidth]{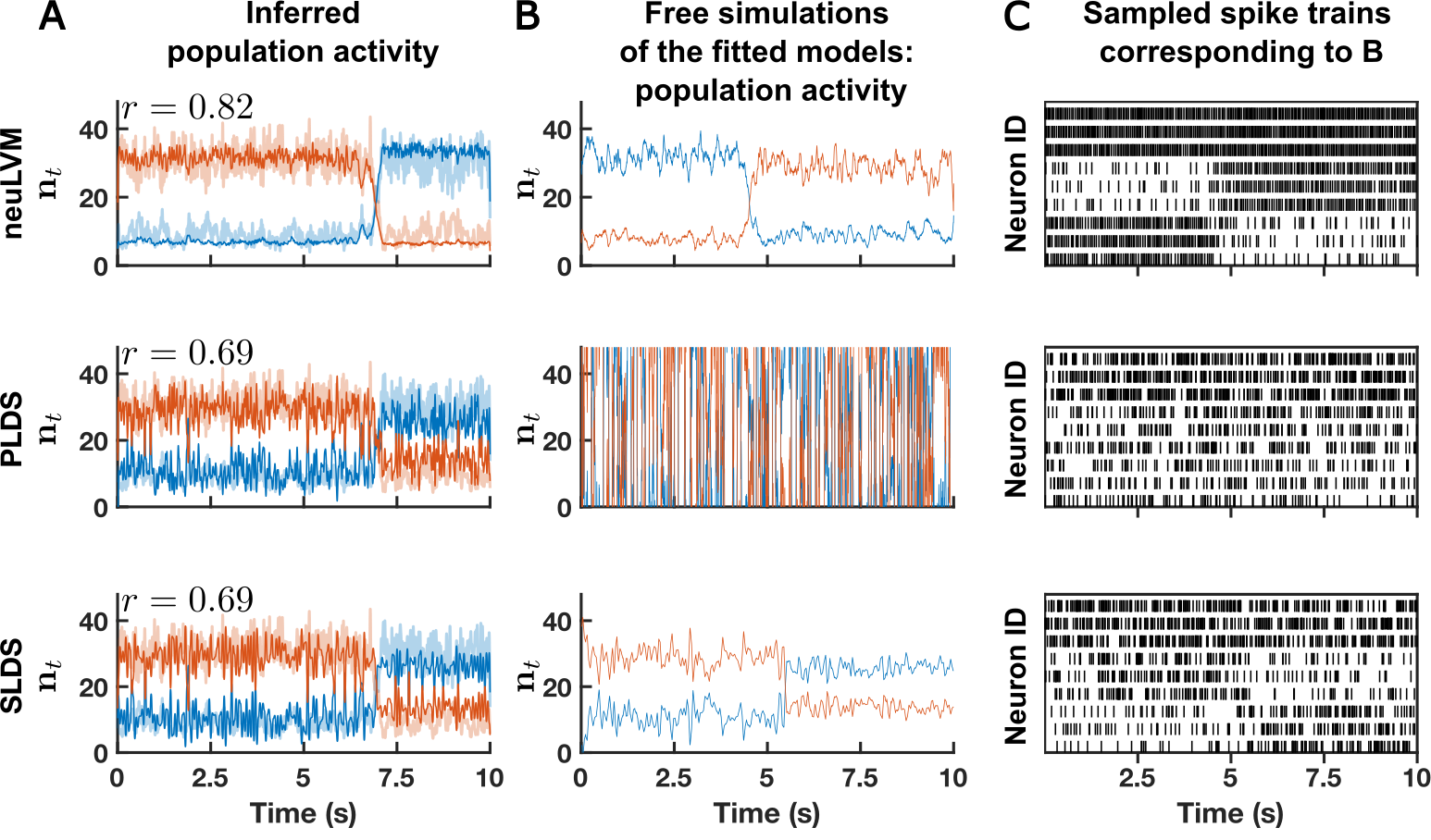}
% \caption{Latent population activity. (A) Pearson correlation coefficient between the inferred population activity and the ground truth. (B) Latent population activity of the two excitatory populations,  for an example trial (the same as Figure~\ref{fig:generated}~A). (C) Population activity generated by free simulations of BILM and PLDS after learning.}\label{fig:latent}
\caption{\textbf{Three-population SNN with metastable point attractors.}
% (A) An example trial generated by the ground truth SNN. The raster plot shows one example training dataset (10 seconds) consisting of spike trains of nine sampled neurons, with the underlying population activity plotted on the right. 
% We find clearly the winner-take-all dynamics with one visible transition.
\textbf{(A)} Latent population activity of the two excitatory populations inferred by neuLVM / PLDS / SLDS for one example trial (the same as in Figure~\ref{fig:3popdata}). The value $r$ is the Pearson correlation coefficient between the inferred $\widehat{\bn}|\byo$ and the ground truth $\bn^*$ population activities.
% The transition period is boxed out, and displayed with higher resolution in Figure~\ref{fig:ap-transition}. 
\textbf{(B-C)} Examples of free simulations of the fitted neuLVM / PLDS / SLDS. 
% (B) An example trail simulated freely by the BILM after training. BILM learns to reproduce the metastable dynamics even when the training data is of small size.
% (C) An example trail simulated freely by the PLDS after training. PLDS fails to capture the dynamics of interest.
% (D) \textcolor{cyan}{An example trail simulated freely by the SLDS after training. SLDS captures the metastable dynamics with limitation: as in the training dataset only the transition from pop.1 to pop.2 is contained, trained SLDS is not able to simulate transitions in the reverse way (pop.2 takes the domination from pop.1).}
% (F) Latent population activity estimated by BILM. \textcolor{cyan}{Different from other methods (including empirical smoothing Figure~\ref{fig:ap-emp}, the inductive bias of BILM enables it to learn to ignore the small fluctuations in stable stages while keeping an accurate account of the transition.}
% (G) Latent population activity estimated by PLDS. The latent trajectory estimated by PLDS cannot be readily interpreted as the population activity, therefore a least squares linear mapping is applied. Even though PLDS fits the winner-take-all dynamics in the training set, it fails to reproduce when simulated freely.
% (H) 
% (D) Pearson correlation coefficient between the inferred population activity and the ground truth. 20 (5, 2) different datasets are trained with size 10 (40, 80) seconds. 
% We find BILM fits population activity significantly better than PLDS and SLDS.
}\label{fig:main-exp}
\end{figure}

In contrast with the experiments of Section~\ref{sec:single} where the neuronal parameters were given, here, neuronal and connectivity parameters are not given to neuLVM (see Appendix~\ref{app:WTA}). We compared the performance of neuLVM with other generative models of spiking data -- PLDS \cite{MacBue12}, SLDS \cite{LinJoh17}, and GLM \cite{Pan04, TruEde05, PilShl08} -- on single trials of the spiking benchmark dataset in two ways: (i) we measured the Pearson correlation $r$ between the inferred latent population activity $\widehat{\bn} | \by^{\mathbf{o}}$ and the ground truth population activity $\bn^{*}$ (Table~\ref{tab:bistable-result}~first column); (ii) we assessed how well could the fitted models reproduce metastable dynamics by counting the occurrences of stochastic switches in free simulations -- or in other words, samples -- of the fitted models (Table~\ref{tab:bistable-result}~second column). Tests (i) and (ii) on an example trial are shown in Figure~\ref{fig:main-exp}.

The Poisson Linear Dynamical Systems approach (PLDS, \cite{MacBue12}) assumes that the recorded spikes can be explained by point processes driven by a latent linear dynamical system of low dimension. The Poisson Switching Linear Dynamical System (SLDS, \cite{AckFu70,GhaHin00,Bar06,FoxSud08,PetYu11,LinJoh17}) extends PLDS by allowing the latent variables to switch randomly between two dynamical systems with distinct parameters. We should stress that, in PLDS and SLDS, the latent variables are \textit{phenomenological} representations of neural population activity which have no direct link with the ground truth population activity $\bn^*$. In order to still make test (i) possible for PLDS and SLDS, we will consider the best linear transformation of the inferred latent variables which minimizes the mean squared error with the ground truth population activity $\bn^*$.
% SLDS, \cite{AckFu70,GhaHin00,Bar06,FoxSud08,PetYu11,LinJoh17}

On test~(i), neuLVM gave better estimates $\widehat{\bn} | \by^{\mathbf{o}}$ of the latent population activity $\bn^*$ (Pearson correlation $r=0.81$) than the best linear transformation of the latent activity inferred by PLDS and SLDS ($r=0.69$ and $r=0.73$ respectively) (Table~\ref{tab:bistable-result}~first column). The GLM approach cannot be included in test (i) since it ignores unobserved neurons. Interestingly, the example trial in Figure~\ref{fig:main-exp}A shows the latent population activity $\widehat{\bn} | \by^{\mathbf{o}}$ inferred by neuLVM is smoother than the ground truth $\bn^*$ before and after the switch (finite-size fluctuations are reduced) but $\widehat{\bn} | \by^{\mathbf{o}}$ and $\bn^*$ closely match around the time of the switch. In contrast, fluctuations are exaggerated for PLDS and SLDS. The population activity estimated by simply summing and smoothing the observed spike trains (Appendix~\ref{app:smoothing}) is shown in Figure~\ref{fig:ap-emp}.

On test~(ii), neuLVM, fitted on a single trial of $10$ seconds, was able to reproduce stochastic switches similar to that of the ground truth SNN (Table~\ref{tab:bistable-result}~second column): free simulations of the fitted neuLVM showed $7.8$ switches in $100$ seconds on average ($10.3$ switches on average for the ground truth SNN). To make sure that stochastic switches were the result of parameter learning via the Baum-Viterbi algorithm, we verified that, before learning, neuLVM did not show any metastable dynamics (Figure~\ref{fig:ap-init}). Examples of simulated trials are shown in Figure~\ref{fig:main-exp}B. PLDS failed to reproduce stochastic switches, which is not surprising since winner-take-all dynamics are typically nonlinear. SLDS could reproduce stochastic switches at the correct mean frequency ($11.9$ instead of the ground truth $10.3$), but the standard deviation of the simulated switch count, $8.9$ ($2.7$ for the ground truth SNN), indicates that a single $10$ seconds trial was probably not sufficient for SLDS to learn switching probabilities reliably. Finally, neuronal stochasticity and small network size ($9$ neurons) did not allow GLM to produce stochastic switches, even when the training trial was prolonged to $500$ seconds.

Taken together, only neuLVM could infer the latent population activity and reliably learn the metastable dynamics on single trials of $10$ seconds, demonstrating the effectiveness of its neuronally-grounded inductive biases. Of course, these results do not guarantee that the inductive biases of neuLVM would be effective on real data since real data is most certainly out-of-distribution for neuLVM. While applications on real data are beyond the scope of this paper, in Appendix~\ref{app:additional_exp}, we show that neuLVM is robust, to a certain extent, to within-population heterogeneity and out-of-distribution data.

\begin{table}[]
\centering
\caption{Model performance summary (corresponding to Figure~\ref{fig:main-exp}).}%.\\
%Standard deviations were computed over models fitted on 20 different trials.}
\label{tab:bistable-result}
\begin{tabular}{ccc}
\toprule
Models       & \begin{tabular}[c]{@{}c@{}}Pearson correlation $r$\\ between $\widehat{\mathbf{n}}|\mathbf{y^o}$ and $\mathbf{n}^*$\end{tabular} & \begin{tabular}[c]{@{}c@{}}Number of switches during $100$ seconds free simulations \\ of the fitted models ($10.3\pm2.7$ for the ground truth SNN)\end{tabular} \\ \midrule
% \textbf{SNN} & \textbf{-}                                                                                                                     & $\mathbf{10.3\pm2.7}$                                                                \\
neuLVM         & $0.81\pm0.02$                                                                                                                  & $7.8\pm4.1$                                                                          \\
PLDS         & ($0.68\pm0.11$)                                                                                                                  & not visible                                                                          \\
SLDS         & ($0.73\pm0.02$)                                                                                                                  & $11.9\pm8.9$                                                                         \\
GLM          & -                                                                                                                              & not visible                                                                          \\ 
\bottomrule
\end{tabular}
Mean and ($\pm$) standard deviation were computed over $20$ different trials. Parentheses for PLDS and SLDS indicate that these results are for the best linear transformation of the inferred latent variables.
\end{table}

\subsubsection{Generalization: towards experimental predictions with neuronally-grounded modeling}
Bottom-up, mechanistic models allow us to perform \textit{in silico} experiments and generate predictions about neural microcircuits, which can then be tested experimentally. So we wondered: can neuLVM, fitted on a single trial of spontaneous activity (like in Section~\ref{sec:reproduction}), predict the response of the SNN when an external perturbation is applied? As a preliminary step in that direction, we tested whether an external stimulation of the fitted model would generate the same response as that of the microscopic SNN when subjected to the same perturbation.

\begin{figure}
\centering
\includegraphics[width=1\linewidth]{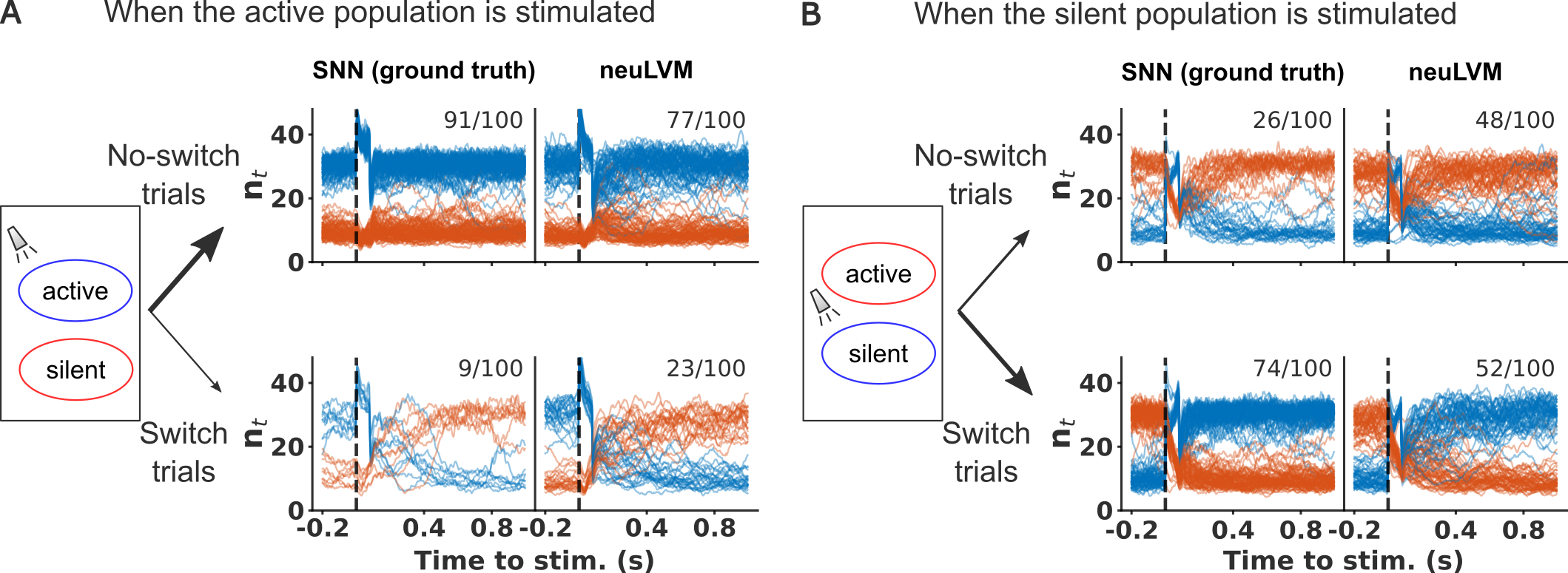}
\caption{\textbf{Network responses to perturbations mimicking optogenetic stimulation.} \textbf{(A)} Activities of the excitatory populations when the active population is stimulated ($100$ trials, ratios indicate the number of No-switch or Switch trials). \textbf{(B)} Same as A but the silent population is stimulated. }\label{fig:photo}
\end{figure}

Using the same multi-population network as in Section~\ref{sec:3pop} (Figure~\ref{fig:3popdata}A) and neuLVM fitted on a single trial of spontaneous activity (Figure~\ref{fig:3popdata}B), we compared the response of the ground truth SNN with that of neuLVM when one of the populations was stimulated by a current pulse of $4$ ms mimicking the stimulation of an optogenetically modified population by a short light pulse. We simulated $100$ trials where the momentarily active excitatory population was stimulated, and 100 where the momentarily silent excitatory population was stimulated (Figure~\ref{fig:photo}A and B respectively). Each stimulation led to two possible outcomes: stimulation could trigger a state switch (Switch trials) or no state switch (No-switch trials).
In both the ground truth SNN and the fitted neuLVM, we found that stimulating the silent population triggered more frequent state switches (Figure~\ref{fig:photo}B) than stimulating the active population (Figure~\ref{fig:photo}A). Moreover, in both the ground truth and the fitted neuLVM, we could induce `excitatory rebound' switches by stimulating the active population (Figure~\ref{fig:photo}A, lower half).

% SNN: ($p=3.3\mathrm{e}{-2}$) 
% neuLVM: ($p=1.1\mathrm{e}{-16}$)
% with probability larger than 5\%

\section{Discussion}

Understanding the neural dynamics underlying computation in the brain is one of the main goals of latent variable modeling of multi-neuronal recordings \cite{ZhaPar16, PanOsh18, DunBoh19, RutBer20, KimLuo21,WhiBut19}. We contribute to this effort by proposing here a bottom-up, mechanistic LVM -- the neuronally-grounded latent variable model (neuLVM) -- which can be mapped to a multi-population SNN. Using SNN-based generative models, which are more biological than RNN-based models \cite{PanOsh18}, could allow systems neuroscientists to test hypotheses about the architecture of the probed microcircuit, and provide a neuronally-grounded understanding of computation. 

While this work shows the potential of the neuLVM approach, the application of neuLVM to real data faces two methodological challenges. First, there is the problem of identifiability: although neuLVM could recover a single unknown connectivity parameter (Section~\ref{sec:single}), our method could not always recover the SNN parameters when many parameters were unknown (Section~\ref{sec:3pop}). Bayesian inference could circumvent the problem of non-identifiability by estimating the full posterior distribution over model parameters \cite{LueGon17,RenLon20}. In addition, perturbing the probed network, with optogenetic stimulation, for example, could help model parameter recovery by providing richer data. Second, in the case of real data, choosing a good generative SNN model is a nontrivial task. For example, how many homogeneous populations should the SNN have? Clustering the recorded spike trains could guide the design of possible generative models and Bayesian model comparison, as used in biophysical modeling of neuroimaging data \cite{PenSte04, PenSte10, BroSch11}, could help in selecting the most likely model among several possible models.

The model proposed here is only one particular example of an SNN-based, tractable latent variable model. Whether other such neuronally-grounded models of partially observed spike trains can be formulated and efficiently applied to real data is a question left for future work. 

\begin{ack}
We thank Johanni Brea for several discussions and for his comments on an early version of this work. We also thank Tilo Schwalger for the discussions and for sharing his code. Code from Joachim Koerfer was also used. This research was supported by Swiss National Science Foundation (no. 200020\_184615).
\end{ack}

%\section*{References}

\medskip
\bibliography{mybib}
%%%%%%%%%%%%%%%%%%%%%%%%%%%%%%%%%%%%%%%%%%%%%%%%%%%%%%%%%%%%
% \begin{comment}

% \end{comment}

%%%%%%%%%%%%%%%%%%%%%%%%%%%%%%%%%%%%%%%%%%%%%%%%%%%%%%%%%%%%

\newpage

\appendix

\newcommand{\toptitlebar}{
  \hrule height 4pt
  \vskip 0.25in
  \vskip -\parskip%
}
\newcommand{\bottomtitlebar}{
  \vskip 0.29in
  \vskip -\parskip
  \hrule height 1pt
  \vskip 0.09in%
}

\newpage
\setcounter{page}{1}
\appendix
\begin{center}
{\Large \textbf{Appendices of:\\}}
\vspace{0.5cm}

\toptitlebar
{\LARGE \textbf{\newtitle}}
\bottomtitlebar
\vspace{0.5cm}
\end{center}

\renewcommand{\thetable}{S\arabic{table}}  
\renewcommand{\thefigure}{S\arabic{figure}}

% \appendix

% \section{Appendix}
\section{Mesoscopic model in the case of LIF neurons}\label{app:meso}
%\begin{equation*}
%    \Lambda_t = \frac{\sum_{a=1}^\mathcal{T}\lambda_{ta}(1 - S_{ta})S_{ta}\,\bn_{t-a}}{\sum_{a=1}^\mathcal{T}(1 - S_{ta})S_{ta}\,\bn_{t-a}}
%\end{equation*}
In this section, we present in detail the mesoscopic model of \citetSM{SchDeg17} in the case of multiple interacting populations of LIF neurons, as formulated in \citeSM{SchLoe21}.

\paragraph{Fine-grained SNN of LIF neurons with escape noise.}
%\paragraph{The probability $\rho^{\Delta t}_{\theta^i}$ in the case of leaky integrate-and-fire neurons with `escape noise' (LIF)}
Let us consider a general network of $N$ LIF neurons (indexed by $i=1, \dots, N$) with escape noise \citeSM{GerKis14}. Neurons are modeled as point processes: the probability for neuron $i$ to emit a spike at time $t$, given the past network activity ${\by}_{1:t-1}$, is
\begin{equation*}
    p({y}^i_{t}=1| {\by}_{1:t-1}, \Theta) = 1-\exp\left(-\lambda^i_{t}\Delta t\right), \quad \text{with } \lambda^i_{t} = \exp\left(V^{i}(t|\hat{t}^i) - \vartheta^i\right),
\end{equation*}
where the escape rate (or stochastic intensity) $\lambda^i_{t}$ depends on the momentary difference between the membrane potential $V^{i}(t|\hat{t}^i)$ and the firing threshold $\vartheta^i$, via an exponential escape function. The voltage $V^{i}(t|\hat{t}^i)$ of neuron $i$ at time $t$ depends on its last spike time $\hat{t}^i = t - a^i$ and the inputs received up to time $t$, which include the inputs coming from the other neurons and the external input $\mathbf{I}^{\mathrm{ext},i}_{1:t}$. Between spikes, for all $t> \hat{t}^i+t^i_{\mathrm{ref}}$ ($t^i_{\mathrm{ref}}$ being the absolute refractory period of neuron $i$), the voltage dynamics follows
\begin{align*}
    V^{i}(t|\hat{t}^i) &= V^{i}(t-1|\hat{t}^i) + \left(\frac{U^i_\mathrm{r} + RI^{\mathrm{ext},i}_t - V^{i}(t-1|\hat{t}^i)}{\tau^i_{\mathrm{mem}}}\right)\Delta t+ 
    \sum_{j=1}^{N}J^{ij}\left(\epsilon^{ij} * \by^j \right)(t),
    % \sum_{\beta=1}^{M}J^{\alpha\beta}\sum_{j=1}^{N^\beta} \left(\epsilon^\beta * s_j^\beta \right)\left(t\right)\right)\Delta t,
\end{align*}
and $V^{i}(t|\hat{t}^i) = 0$, for all $t\leq \hat{t}^i+t^i_{\mathrm{ref}}$ (which means that the voltage is reset to $0$ after each spike and is clamped at $0$ for an absolute refractory period $t^i_{\mathrm{ref}}\geq 0$). The parameters $\tau^i_{\mathrm{mem}}>0$ and $U^i_\mathrm{r}>0$ are the membrane time constant and the resting potential respectively. The neuron $i$ is therefore characterized by the parameters $\theta^i = \{\vartheta^i, U^i_{\mathrm{r}}, \tau^i_{\mathrm{mem}}, t^i_{\mathrm{ref}}\}$. While the escape function is usually parameterized by a rescaled exponential function of the form $f(v) = \frac{1}{\tau^i_0}\exp(\beta^i(v-\tilde{\vartheta}^i))$ \citeSM[Sec~9.1]{GerKis14}, the parameters $\tau^i_0, \beta^i$ and $\tilde{\vartheta}^i$ can be absorbed in $\vartheta^i$ (up to a rescaling of the resting potential $U^i_{\mathrm{r}}$). The resistance $R=1\,\Omega$ is used here simply for the consistency of physical units. The postsynaptic current induced by a spike of neuron $j$ on neuron $i$ is defined by the synaptic weight $J^{ij}$ and the synaptic kernel $\epsilon^{ij}:\R_+\to\R_+$. In this work, we consider exponential kernels of the form $\epsilon^{ij}(t) = \frac{\mathcal{H}(t-\Delta^{ij})}{\tau^{ij}_{\mathrm{syn}}}\exp\left(-\frac{t-\Delta^{ij}}{\tau^{ij}_{\mathrm{syn}}}\right)$, where $\tau^{ij}_{\mathrm{syn}}$ is the synaptic time constant, $\Delta^{ij}$ is the synaptic delay and $\mathcal H$ is the Heaviside function. The symbol $\ast$ denotes the convolution operator. 

%$J^{ij} \in \mathbb{R}$ is the connection strength from the neuron $j$ to the neuron $i$. The synaptic kernel $\epsilon^j\left(t\right)$ is defined as the normalized postsynaptic current (PSC) induced by one input spike from the neuron $j$. In this paper, we consider $\epsilon^j\left(t\right)$ as the exponential filter $\frac{1}{\tau^j_{\mathrm{syn}}}\exp\left(-\frac{[t-\Delta^j]_+}{\tau^j_{\mathrm{syn}}}\right)$ with $\Delta^j$ being the synaptic delay. $\by^j\in\{0,1\}^{T}$ is the spike train of neuron $j$.

%\paragraph{The probability $\rho^{\Delta t}_{\theta}$ in the case of a homogeneous population of LIF neurons}
\paragraph{Coarse-grained multi-population SNN.}\label{app:coarse-grained-SNN}
Coarse-graining and mean-field approximations consist in partitioning the $N$ neurons into $K$ homogeneous populations, indexed by $\alpha = 1, \dots, K$, where (i) all the neurons $i$ in population $\alpha$ share the same neuronal parameters $\theta^i=\theta^\alpha$; (ii) for any neuron $j$ in population $\beta$ and any neuron $i$ in population $\alpha$, $J^{ij}=J^{\alpha\beta}/N^\beta$ ($N^\beta$ being the number of neurons in population $\beta$) and $\epsilon^{ij}= \epsilon^{\alpha\beta}$; (iii) all the neurons $i$ in population $\alpha$ share the same external input $\mathbf{I}^{\mathrm{ext},i} = \mathbf{I}^{\mathrm{ext},\alpha}$. In such a coarse-grained $K$-population SNN, we have, for any neuron $i$ in population~$\alpha$,
\begin{equation*}
    \sum_{j=1}^{N}J^{ij}\left(\epsilon^{ij} * \by^j \right)(t) = \sum_{\beta=1}^{K}J^{\alpha\beta}\left(\epsilon^{\alpha\beta} * \bn^\beta \right)(t)/N^{\beta},
\end{equation*}
where $n^\beta_{t} = \sum_{i\,\in\, \text{pop. $\beta$}} y^i_{t}$ is the total number of spikes in population~$\beta$ at time $t$. Hence, the probability for \textit{any} neuron $i$ in population~$\alpha$ to emit a spike at time $t$, given its age $a$ and the past population activity $\bn_{1:t-1}$ is
\begin{equation}\label{eq:pfire_multi}
    p_{t,a}^{\mathrm{fire}, \alpha} = 1-\exp\left(-\lambda^\alpha_{t}\Delta t\right), \quad \text{with } \lambda^\alpha_{t} = \exp\left(V^{\alpha}(t|t-a) - \vartheta^\alpha\right). \\
\end{equation}
For all $a> t^\alpha_{\mathrm{ref}}$, we have the update rule
\begin{multline*}
    V^{\alpha}(t|t-a) = V^{\alpha}(t-1|t-a) + \left(\frac{U^\alpha_\mathrm{r} + RI^{\mathrm{ext},\alpha}_t - V^{\alpha}(t-1|t-a)}{\tau^\alpha_{\mathrm{mem}}}\right)\Delta t \\
    + \sum_{\beta=1}^{K}J^{\alpha\beta}\left(\epsilon^{\alpha\beta} * \bn^\beta \right)(t)/N^{\beta},
    % \sum_{\beta=1}^{M}J^{\alpha\beta}\sum_{j=1}^{N^\beta} \left(\epsilon^\beta * s_j^\beta \right)\left(t\right)\right)\Delta t,
\end{multline*}
and $V^{\alpha}(t|t-a) = 0$ for all $a \leq t^\alpha_{\mathrm{ref}}$. This gives the explicit expression for the probability $p_{t,a}^{\mathrm{fire}}$ in Eq.~\eqref{eq:neuron-model}. In this work, for simplicity, we will assume that all the synaptic kernels are the same, i.e. $\epsilon^{\alpha\beta}=\epsilon$, $\forall \alpha,\beta$ (see Table~\ref{tab:gt-parameters}).

%In the case where all the $N$ LIF neurons mentioned above are of one homogeneous population, all the connection strength are identical, that is, $J^{ij}=J/N$ (mean-field approximation) and all the neurons share the same parameters, that is, $\theta^i=\theta$ where $\theta=\{\tau_{\mathrm{mem}},\vartheta,U_\mathrm{r},t_{\mathrm{ref}}, \tau_{\mathrm{syn}},\Delta\}$.  

\paragraph{Mesoscopic description.} The $K$-population SNN described above does not by itself constitute a mesoscopic model because the probability $p_{t,a}^{\mathrm{fire}, \alpha}$ still involves the age $a$ of some neuron. To get a mesoscopic model (i.e. a model that does not involve the fine-grained modeling of each individual neuron), \citetSM{SchDeg17} used the population activity $\bn$ to approximate the age density of each population and derived a closed-form system of stochastic integral equations: For all $\alpha \in 1, \dots, K$,
\begin{subequations}\label{eq:stochastic_integral_multi}
\begin{align}
    n^\alpha_t &\sim \text{Binomial}\left(N^\alpha,\; \bar{n}^\alpha_t /N^\alpha\right),\label{eq:binomial_multi}\\
    \bar{n}^\alpha_t &= \Bigg[\sum_{a\geq 1}p^{\mathrm{fire}, \alpha}_{t,a}S^\alpha_{t,a}\,n^\alpha_{t-a} + \Lambda^\alpha_t\bigg(N^\alpha - \sum_{a \geq 1}S^\alpha_{t,a}\,n^\alpha_{t-a}\bigg)\Bigg]_+, \\
    \Lambda^\alpha_t &= \frac{\sum_{a\geq 1}p^{\mathrm{fire}, \alpha}_{t,a}(1 - S^\alpha_{t,a} )S^\alpha_{t,a}\,n^\alpha_{t-a}}{\sum_{a\geq 1}(1 - S^\alpha_{t,a})S^\alpha_{t,a}\,n^\alpha_{t-a}},
\end{align}
\end{subequations}
where $S^\alpha_{t,a}=\prod_{s=0}^{a-1}(1 - p^{\mathrm{fire}, \alpha}_{t-a+s,s})$ is the survival, i.e. the probability for a neuron in population~$\alpha$ to stay silent between time $t-a$ and $t-1$. A concise version of the derivation of the mesoscopic model~\eqref{eq:stochastic_integral_multi} is presented in \citeSM{SchLoe21}. Note that Eq.~\eqref{eq:stochastic_integral_multi} is not a one-dimensional stochastic dynamical system: the Markov embedding of the stochastic dynamics~\eqref{eq:stochastic_integral_multi} is infinite-dimensional \citeSM{SchLoe21}. Indeed, Eq.~\eqref{eq:stochastic_integral_multi} does not only describes the evolution of the population activity $n_t^\alpha$ but it also describes the evolution of the whole age (pseudo) density $\{S_{t,a}^\alpha n_{t-a}\}_{a\geq 1}$ in the population, also called the ``refractory density'' \citeSM{SchChi19}.
%\paragraph{The `modulating factor' $\Lambda_t$}
%In a homogeneous population, $\sum_{j=1}^{N}J^{ij}\left(\epsilon^j * \by^j \right)\left(t\right)$ could be rewritten as $J\left( \epsilon * n_t  \right)\left(t\right)/N$ where $n_{t} = \sum_{i=1}^N y^i_{t}$ being the total number of spikes in the population at time $t$. 
%Further, for \textit{any} neuron in the population with the same last spike time (i.e., the same age $a$), they share the same probability to emit a spike ($p_{t,a}^{\mathrm{fire}}$) at time $t$. Hence, the probability for any neuron $i$ in population $\alpha$ to to emit is spike at time $t$

Formally, the `initial condition' of Eq.~\eqref{eq:stochastic_integral_multi} is defined by the population activity $\bn_t$ for all $t\leq 0$ (denoted $\bn_{t\leq 0}$). Several practical choices of initial conditions have been discussed in \citeSM{SchDeg17,RenLon20,SchLoe21}. In this work, if not otherwise specified, $\bn_{t\leq 0}$ is taken to be time-invariant, with stationary activities estimated from the observed data (see below).

The size of the discrete time steps $\Delta t$ does not need to be the same for the fine-grained SNN and for the mesoscopic model~\eqref{eq:stochastic_integral_multi}. Indeed, it can be useful to take longer time steps for the mesoscopic description (time coarse-graining). In the following appendices, when there is an ambiguity, $\Delta t_{\mathrm{meso}}$ will denote the time step length for the mesoscopic model and neuLVM. The length $\Delta t_{\mathrm{meso}}$ will always be smaller or equal to the neuronal absolute refractory periods so that a neuron can fire at most once in each time step.

\section{neuLVM for multiple interacting populations}\label{app:neuLVM}
Let us assume that we observe, during $T$ time steps, the spike trains of $q$ simultaneously recorded neurons that are part of a $K$-population SNN of $N$ neurons, with $N>q$. For each of the population $\alpha=1,\dots,K$, $q^\alpha>0$ neurons are observed $(\sum_{\alpha=1}^Kq^\alpha=q)$ and share the same set of neuronal parameters $\theta^\alpha$, input weights $\{J^{\alpha\beta}/N^\beta\}_{\beta=1}^K$, and output weights $\{J^{\beta\alpha}/N^\alpha\}_{\beta=1}^K$, where $N^1, \dots, N^K$ are the numbers of neurons in each population $(\sum_{\alpha=1}^K N^\alpha = N)$. 
%$\Theta^\alpha=\{J^{\alpha\beta}, \theta^\alpha\}$.

\paragraph{The likelihood $\LL$ of the observed spike trains.} Following the assumptions described above, the likelihood $\LL$ of the observed spike trains $\byo$ (a binary $q\times T$ matrix) can be formally written as $\sum_\bn p(\byo,\bn| \Theta)$, where $\bn$ (an integer-valued $K\times T$ matrix) is the population activity and $\Theta = \{\{J^{\alpha\beta}\}_{1\leq \alpha, \beta \leq K}, \{\theta^\alpha\}_{\alpha=1}^K\}$ are the parameters of the $K$-population SNN. The probability $p(\byo,\bn| \Theta)$ factorizes in $T$ terms of the form
\begin{equation*}
      p(\by^{\mathbf{o}}_t, \bn_t | \by^{\mathbf{o}}_{1:t-1}, \bn_{1:t-1}, \Theta)  = \underbrace{p(\by^{\mathbf{o}}_t|\by^{\mathbf{o}}_{1:t-1}, \bn_{1:t-1}, \Theta)}_{\text{\textbf{part a}}}\underbrace{p(\bn_t|\bn_{1:t-1}, \Theta)}_{\text{\textbf{part b}}}.   
\end{equation*}

The probability (\textbf{part a}) of the observed spikes $\by^{\mathbf{o}}_t$ at time $t$ given the past observed spike activity $\by^{\mathbf{o}}_{1:t-1}$ and the past population activity $\bn_{1:t-1}$ is

\begin{equation*}
    p(\by^{\mathbf{o}}_t|\by^{\mathbf{o}}_{1:t-1}, \bn_{1:t-1}, \Theta)
    =\prod_{\alpha=1}^K \prod_{i=1}^{q^\alpha} p(y^{\mathbf{o},\alpha,i}_t|a^i, \bn_{1:t-1}, \Theta)
    =\prod_{\alpha=1}^K \prod_{i=1}^{q^\alpha} p_{t,a^i}^{\mathrm{fire}, \alpha},
\end{equation*}
where $p_{t,a^i}^{\mathrm{fire}, \alpha}$, given by Eq.~\eqref{eq:pfire_multi} in Appendix~\ref{app:coarse-grained-SNN}, is the probability for the recorded neuron $i$ of population $\alpha$ to emit a spike at time $t$.

The probability (\textbf{part b}) of the population activity $\bn_t$ at time $t$ given the past population activity $\bn_{1:t-1}$ is
\begin{equation*}
    p(\bn_t|\bn_{1:t-1},\Theta)
    =\prod_{\alpha=1}^K p(n^\alpha_t|\bn_{1:t-1}, \Theta),
\end{equation*}

where $p(n^\alpha_t|\bn_{1:t-1}, \Theta)$ is approximated by the mesoscopic model~\eqref{eq:stochastic_integral_multi}.  

\section{Fitting algorithm for neuLVM}\label{app:neuLVM_fit_appendix}

\paragraph{Baum-Viterbi algorithm.} Given the observed spike trains $\byo$, we optimize the likelihood $\LL=\sum_\bn p(\byo,\bn| \Theta)$ via an EM-like algorithm -- the Baum-Viterbi algorithm \citeSM{EphMer02}. 
Relying on the heuristic that the posterior $p(\bn|\byo,\Theta)$ should be concentrated around its maximum, we approximate the posterior $p(\bn|\byo,\Theta)$ by a point mass $\delta_\mu$, where $\mu=\argmax_{\bn}\log p(\byo, \bn|\Theta)$. 
By doing so, the alternating estimation (E) and maximization (M) step of the $n$-th iteration read
\begin{enumerate}
    \item[\textbf{E-step}.] $\widehat{\bn}^{n} = \operatorname{argmax}_\bn \log p(\byo,\bn| \widehat{\Theta}^{n-1})$,
    % \item[\textbf{Step E}.] Optimize $\bn$ in $\log p(\byo,\bn| \widehat{\Theta}^{n})$ with gradient ascent. Write $\widehat{\bn}^{n+1}$ the result of this optimization step
    \item[\textbf{M-step}.] $\widehat{\Theta}^{n} = \operatorname{argmax}_\Theta \log p(\byo,\widehat{\bn}^{n}| \Theta)$.
    % \item[\textbf{Step M}.] Optimize $\Theta$ in 
    %     $\log p(\byo,\widehat{\bn}^{n+1}| \Theta)$ 
    %     with gradient ascent. Write $\widehat{\Theta}^{n+1}$ the result of this optimization step.
\end{enumerate}
\paragraph{Details of the optimization.} 
In the \textbf{M-step}, parameters $\Theta$ are optimized using the L-BFGS-B algorithm and the optimization stops when either the maximum number of iterations ($\text{maxiter}_{\text{M}}$) is reached, or the objective function improves by less than $\text{ftol}_{\text{M}}$, or the maximum norm of the gradient is less than $\text{gtol}_{\text{M}}$. Hyper-parameters including $\text{maxiter}_{\text{M}}$, $\text{ftol}_{\text{M}}$ and $\text{gtol}_{\text{M}}$ are given in Table~\ref{tab:hyper-fitting}.
In the \textbf{E-step}, to carry out gradient ascent, we approximate the discrete Binomial distribution Eq.~\eqref{eq:binomial_multi} by a Gaussian, i.e. $n_t^\alpha \sim \mathcal{N}\left(\bar{n}_t^\alpha, \bar{n}_t^\alpha\right)$, where $\bar{n}_t^\alpha$ is given by the mesoscopic model Eq~\eqref{eq:stochastic_integral_multi} \citeSM{SchDeg17}.
% With this approximation, we get the differentiable expression
% \begin{equation*}
%     \log p(\bn|\Theta) \propto -\frac{1}{2}\sum_{t=1}^T \left\{\log(\bar{n}_t) + \frac{(\bn_t - \bar{n}_t)^2}{\bar{n}_t}\right\}. 
% \end{equation*}
With this approximation, the latent population activity $\bn$ is optimized with the Adam algorithm with learning rate $\text{lr}_{\text{E}}$ and the optimization stops when either the maximum number of iterations ($\text{maxiter}_{\text{E}}$) is reached, or the objective function stops improving for the last $\text{itertol}_{\text{E}}$ iterations. Hyper-parameters including $\text{lr}_{\text{E}}$, $\text{maxiter}_{\text{E}}$ and $\text{itertol}_{\text{E}}$ are given in Table~\ref{tab:hyper-fitting}.
The estimated parameters $\widehat{\Theta}$ and the estimated latent population activity $\widehat{\bn}$ are the result of many iterations of \textbf{E-step} and \textbf{M-step}. The fitting algorithm ends either when it stops improving the objective function or the maximum number of E-M iterations is reached.

\paragraph{Multiple data-driven initializations.} To deal with the fact that the joint probability $p(\byo,\bn | \Theta)$ to optimize is non-convex and high-dimensional ($\bn$ has dimension $K\times T$), we perform the Baum-Viterbi algorithm $\text{N}_{\text{init}}$ times with initial parameters $\widehat{\Theta}^0$ uniformly sampled in a certain range given in Appendices~\ref{app:1pop} and \ref{app:WTA}. Since the sum over the observed neurons from population $\alpha$, $\sum_{i=1}^{q^\alpha} \byoi_{1:T}$, already provides a rough estimate of the latent population activity $\bn_{1:T}^\alpha$, the \textbf{E-Step} of the first iteration ($\widehat{\bn}^1$) is replaced by an empirical estimation of the population activity $\widehat{\bn}^{\mathrm{sm}}_\sigma$ from the observed spike trains (see Appendix~\ref{app:smoothing}).
%the sum over the observed neurons (for each population), smoothed with a Gaussian kernel with standard deviation $\sigma$. The hyper-parameters $\sigma$ are given in Appendices~\ref{app:1pop} and \ref{app:WTA}.
%\begin{itemize}
    %\item Since the the sum over the observed neurons from population $\alpha$, $\sum_{i=1}^{q^\alpha} \byoi_{1:T}$, already provides a rough estimate of the latent population activity $\bn_{1:T}^\alpha$, we replace the \textbf{E-Step} of the first iteration ($\widehat{\bn}^1$) by the sums over the observed neurons (for each population), smoothed with a Gaussian kernel with standard deviation $\sigma$. The hyper-parameter $\sigma$ is given in Appendices~\ref{app:1pop} and \ref{app:WTA}.
    % Take a collection of smoothing width $\{\sigma_1, \dots, \sigma_L\}$ as candidates and select the set of estimates with the largest joint probability $\log p(\byo,\widehat{\bn}| \widehat{\Theta})$. 
    %\item 
%\end{itemize}

\paragraph{Numerical implementation of the mesoscopic model.} To implement the mesoscopic model~\eqref{eq:stochastic_integral_multi}, we approximate the infinite sums $\sum_{a\geq 1}$ in Eq.~\eqref{eq:stochastic_integral_multi} by finite sums $\sum_{a=1}^{a_{\mathrm{max}}}$, where $a_{\mathrm{max}}$ is chosen to be large enough such that the probability for a neuron to remain silent for a duration longer than $a_{\mathrm{max}}$ is negligible. In our numerical implementation, the mesoscopic model~\eqref{eq:stochastic_integral_multi} has therefore a finite memory $a_{\mathrm{max}}$. Note that a more principled way to implement finite memory can be found in \citeSM{SchLoe21}, where a numerical implementation similar to ours is presented in detail. The hyper-parameter $a_{\mathrm{max}}$ is given in  Appendices~\ref{app:1pop} and \ref{app:WTA}. If not otherwise specified, the initial condition $\bn_{\leq 0}$ of Eq.~\eqref{eq:stochastic_integral_multi} are chosen to be time-invariant, with stationary activities estimated from the first $a_{\mathrm{max}}$ time steps of the recorded spike trains.

\section{Smoothed empirical population activity}\label{app:smoothing}

A smoothed empirical estimation of the population activity $\widehat{\bn}^{\mathrm{sm}}_{\sigma}$ was obtained from the recorded spike trains $\byo$ by applying a Gaussian smoothing kernel $g_\sigma$ with standard deviation $\sigma$. For population $\alpha=1,\dots,K$,
\begin{equation*}
      \widehat{\bn}^{\mathrm{sm},\alpha}_{\sigma,t}  = \left(\left(\frac{N^\alpha}{q^\alpha}\sum_{i=1}^{q^\alpha}\by^{\mathbf{o},\alpha,i}_{1:T}\right)*g_\sigma\right)(t).
\end{equation*}
%The symbol * denotes the convolution operator.

\section{Details of the cluster state example}\label{app:1pop}

Values of parameters used in this example are given in Table~\ref{tab:gt-parameters}, except if mentioned otherwise.

\paragraph{When the network is initialized in the unstable asynchronous state (Figure~\ref{fig:1pop-exp}B,C).} 
In this case, the network is always initialized, at time 0, in the same unstable asynchronous state with a firing rate of 20 Hz.
The spike train power spectra (Figure~\ref{fig:1pop-exp}B), for different choices of connectivity parameter $J$, were computed using 600 non-overlapping segments of $120$ s.
To measure the goodness of the connectivity recovered by newLVM, for each $J$ in $\{59,60,61,62,63,64,65\}$ mV, we simulated the ground truth SNN (starting from the same unstable asynchronous state mentioned above) for $1$ s and further generated $10$ different datasets with different samples of six observed neurons ($1\%$ of the population). 
% The neuLVM was fitted $10$ times (one fit each dataset, each with a different initial $\widehat{J}^0$) for each $J$ and the learned $\widehat{J}$s were compared to the ground truth $J$s in Figure~\ref{fig:1pop-exp}~C.

\paragraph{When the network is initialized in a 4-cluster state (Figure~\ref{fig:1pop-exp}D-F).} 
In this case, we simulated a trial (one second, $J=60.32$ mV) with a transition from a metastable 4-cluster state to a 3-cluster state (Figure~\ref{fig:1pop-exp}D,E). 
To test how well newLVM work in the regime where only a tiny fraction of the total number of neurons is observed, for each number $\{1,2,5,10\}$ of observed neurons, we generated 10 different datasets with different samples of observed neurons.  

\paragraph{Fitting of the neuLVM.} 
The initial parameter $\widehat{J}^{0}$ was drawn uniformly in $[10,30]\cup[90,110]$ mV. The latent population activity was initialized as the smoothed empirical population activity
($\widehat{\bn}^1=\widehat{\bn}^{\mathrm{sm}}_{\sigma,t}$, Appendix~\ref{app:smoothing}) with $\sigma=1.4$ ms (Figure~\ref{fig:1pop-exp}E). Since the Baum-Viterbi algorithm converged reliably when only $J$ was unknown, $\text{N}_{\text{init}}$ was set to $1$. The hyper-parameter $\Delta t_{\mathrm{meso}}$ was set to $1$ ms and $a_{\mathrm{max}}$ was set to $100$ ($a_{\mathrm{max}}\Delta t_{\mathrm{meso}}=100$ ms).

\paragraph{Fitting of René et al. (2020).}\label{app:Rene}
A naive application of \citetSM{RenLon20} consists in fitting the model with $\widehat{J}=\operatorname{argmax}_\Theta \log p(\widehat{\bn}^{1}| J)$.
The parameter $\widehat{J}^{0}$ and the latent population activity $\widehat{\bn}^1$ were set the same way as for neuLVM, but $\text{N}_{\text{init}}$ was set to $200$. The best performing $\widehat{J}$s were reported in Figure~\ref{fig:1pop-exp}F. The hyper-parameters $\Delta t_{\mathrm{meso}}$ and $a_{\mathrm{max}}$ were the same as for neuLVM. 

\section{Details of the metastable point attractors example}\label{app:WTA}
Values of parameters used in this example are given in Table~\ref{tab:gt-parameters}, except if mentioned otherwise.  
In this example, we simulated a $500$ s-long trial and randomly cut out $20$ non-overlapping $10$ s-segments to generate the training datasets. 

\begin{figure}
\centering
\includegraphics[width=0.35\linewidth]{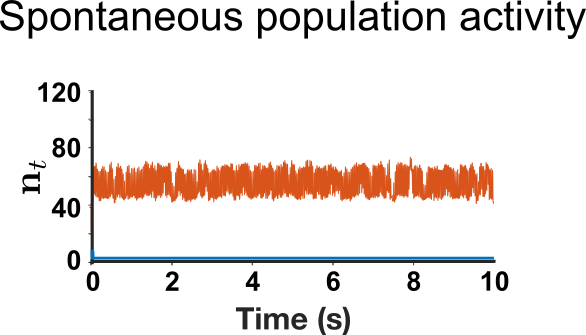}
\caption{Spontaneous population activity simulated by the neuLVM before learning. Population activity of one excitatory population (the blue trace) quickly dies out. No visible metastable dynamics.}\label{fig:ap-init}
\end{figure}

\paragraph{Fitting of the neuLVM.} 
The initial parameters $\widehat{\Theta}^0$ (which include the connectivities $J^{\cdot}$, membrane time constants $\tau_{\mathrm{mem}}^{\cdot}$, firing thresholds $\vartheta^{\cdot}$ and resting potentials $U_\mathrm{r}^{\cdot}$) were sampled randomly by assuming the uniform prior on the range $0.4$ to $2$ times the ground truth values. In this example, the connectivity matrix $\mathbf{J}$ was parametrized by $\{J^{\textrm{e}^1} J^{\textrm{e}^2}, J^{\textrm{i}}\}_{J>0}$: $\mathbf{J}=
%\bigg(\begin{smallmatrix}
%  J^{\textrm{e}^1\textrm{e}^1} & J^{\textrm{e}^2\textrm{e}^1} & J^{\textrm{i}\textrm{e}^1}\\
%  J^{\textrm{e}^1\textrm{e}^2} & J^{\textrm{e}^2\textrm{e}^2} & J^{\textrm{i}\textrm{e}^2}\\
%  J^{\textrm{e}^1\textrm{i}} & J^{\textrm{e}^2\textrm{i}} & J^{\textrm{i}\textrm{i}}\\
%\end{smallmatrix}\bigg)
\bigg(\begin{smallmatrix}
  J^{\textrm{e}^1} & 0 & 0\\
  0 & J^{\textrm{e}^2} & 0\\
  0 & 0 & J^{\textrm{i}}\\
\end{smallmatrix}\bigg)\bigg(\begin{smallmatrix}
  1 & 0 & 1\\
  0 & 1 & 1\\
  -1 & -1 & -1\\
\end{smallmatrix}\bigg)$ (see Figure~\ref{fig:3popdata}A for the network architecture).
The latent population activity was initialized as the smoothed empirical population activity ($\widehat{\bn}^1=\widehat{\bn}^{\mathrm{sm},\alpha}_{\sigma,t}$, Appendix~\ref{app:smoothing}) with $\sigma=400$ms (Figure~\ref{fig:ap-init}).
% The fitting algorithm worked reliably when only $J$ was unknown. 
Out of $5$ fits ($\text{N}_{\text{init}}=5$), the fit with the highest joint likelihood $p(\byo, \widehat{\bn}|\widehat{\Theta})$ was selected. 
% As the empirically smoothed $\bn^{1}$ already provided a good estimation of the ground truth population activity (Figure~\ref{fig:ap-emp}), single \textbf{M-step} was sufficient to fit the model which could reproduce the metastable dynamics. Therefore, in this example, only one iteration of \textbf{M-step} was carried out. Additional \textbf{E-step}s were included to fine-tune the inferred population activity. 
The related hyper-parameter $\Delta t_{\mathrm{meso}}$ was set to $4$ ms and $a_{\mathrm{max}}$ was set to $250$ ($a_{\mathrm{max}}\Delta t_{\mathrm{meso}}=1000$ ms). When $\Delta t_{\mathrm{meso}}$ was set to a value that was larger than the $\Delta t$ of the recorded data, the recorded spike trains were downsampled.

\paragraph{PLDS}\label{app:PLDS}
We used code from \url{https://bitbucket.org/mackelab/pop_spike_dyn/src/master/}.
To fit Poisson Linear Dynamical Systems (PLDS) \citeSM{MacBue12} to the three-population example, we initialized the parameters with nuclear norm penalized rate estimation \citeSM{PfaPne13} and %updated the parameters with Laplace inference.
used the variational EM algorithm of \citeSM{MacBue12}.
The dimensionality of the latent states was set to three (the number of populations). The time resolution of the recorded spike trains was downsampled to $4$ ms ($\Delta t_{\text{PLDS}}=4$ ms). Other hyper-parameters were set to default.

\paragraph{SLDS}\label{app:SLDS}
We used code from \url{https://github.com/lindermanlab/ssm} \citeSM{LinJoh17}.
To fit Poisson Switching Linear Dynamical Systems (SLDS) \citeSM{AckFu70,GhaHin00,Bar06,FoxSud08,PetYu11} to the three-population example, we updated the parameters with stochastic variational inference with the posterior approximated by a factorized distribution.
The dimensionality of the continuous latent states was chosen to be three (the number of populations) and the dimensionality of the discrete latent states was chosen to be three (corresponding to the number of metastable states plus one for the transition state). We specified the `emissions model' as `Poisson\_orthog' with the exponential escape function. Other hyper-parameters were set to default.
Further, for SLDS to work, the discrete time step had to be large enough. Here we downsampled datasets to $40$ ms (the smallest $\Delta t_{\text{SLDS}}$ that worked).

\paragraph{An additional test.}\label{app:additional_exp} We were interested to find out whether neuLVM is robust to within-population heterogeneity and slightly out-of-distribution data. To answer this question, we performed an additional test where we introduced within-population heterogeneity in the ground truth winner-take-all (WTA) network (Section~\ref{sec:3pop}) by adding noise to the connectivity and neuronal parameters as specified in the Table~\ref{tab:3pop-result-hetero} (noise in the neuronal parameters is small to conserve metastable WTA dynamics). 
Furthermore, we set the $N$'s of the neuLVM to 300, 300, 300 (the $N$'s of the ground truth network are 400, 400, 200). We tested neuLVM on eight 10 s-segments cut out from a 100 s-long trial. The method is only mildly affected by these changes: all fitted neuLVM reproduced metastable WTA dynamics and the Pearson correlation between $\widehat{\bn}|\byo$ and $\bn^*$ was $0.76\pm0.02$, which is still higher than the correlations obtained by PLDS and SLDS (see Table~\ref{tab:bistable-result}).

% \begin{figure}
% \centering
% \includegraphics[width=0.25\linewidth]{Figures/Appendix_GLM.png}
% \caption{}\label{fig:ap-GLM}
% \end{figure}

\begin{figure}
\centering
\includegraphics[width=0.3\linewidth]{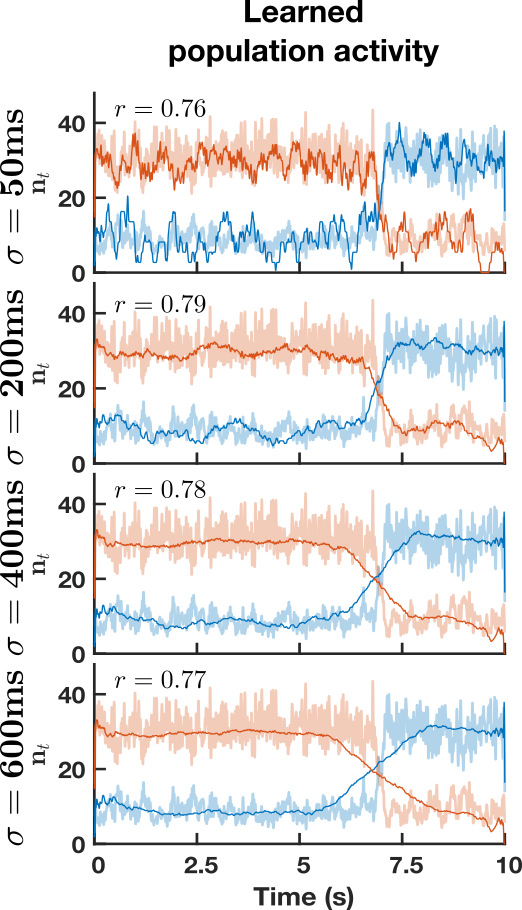}
\caption{Smoothed empirical estimate $\widehat{\bn}^{\mathrm{sm},\alpha}_{\sigma,t}$ (Appendix~\ref{app:smoothing}) of the latent population activity for one example trial (the same as in Figure~\ref{fig:3popdata}, two excitatory populations). The value $r$ is the Pearson correlation coefficient between the inferred $\widehat{\bn}|\byo$ and the ground truth $\bn^*$ population activities.}
\label{fig:ap-emp}
\end{figure}

\begin{table}[]
% \centering
\begin{center}
\caption{Values of parameters used in simulations. \textbf{Boldface} is used to indicate fitted parameters.}
\label{tab:gt-parameters}
\begin{tabular}{lllll}
\toprule
\multicolumn{1}{c}{} &Name & Description              & \multicolumn{2}{c}{Value}                                                                                                                                                                       \\ \cmidrule(r){4-5} 
\multicolumn{1}{c}{} &\multicolumn{1}{c}{}     & \multicolumn{1}{c}{}                         & \multicolumn{1}{c}{\begin{tabular}[c]{@{}c@{}}Example Section~\ref{sec:single}\\ Single excit. population\end{tabular}} & \multicolumn{1}{c}{\begin{tabular}[c]{@{}c@{}}Example Section~\ref{sec:3pop}\\ Excitat. (inhib.) populations\end{tabular}} \\ \midrule
\multicolumn{1}{c}{} & $\Delta t$               & time step                                    & 1\,ms                                                                                      & 0.2\,ms                                                                                                \\
\multicolumn{1}{c}{} & $N$                        & number of neurons                            & 600                                                                                      & 400 (200)                                                                                            \\ \midrule
$\Theta$ & \textbf{$J$}             & \textbf{connectivity} & \textbf{60.32\,mV}                                                                           & \textbf{9.984\,mV (-19.968\,mV)}*                                                                             \\ 
\multicolumn{1}{c}{}& \textbf{$\vartheta$}        & \textbf{firing threshold}                    & 49.7\,mV                                                                          & \textbf{3.7\,mV (3.7\,mV)}                                                                               \\
\multicolumn{1}{c}{}& \textbf{$U_\mathrm{r}$}           & \textbf{resting potential}                   & 26\,mV                                                                            & \textbf{14.4\,mV (14.4\,mV)}                                                                             \\
\multicolumn{1}{c}{}& \textbf{$\tau_{\mathrm{mem}}$}        & \textbf{membrane time constant}              & 100\,ms                                                                           & \textbf{20\,ms (20\,ms)}                                                                                 \\

\multicolumn{1}{c}{}& $t_{\mathrm{ref}}$                & absolute refractory period                   & 0\,ms                                                                                      & 4\,ms (4\,ms)                                                                                            \\ \midrule
$\epsilon$ &$\tau_{\mathrm{syn}}$             & synaptic time constant                       & 4\,ms                                                                                      & 3\,ms (6\,ms)                                                                                            \\
\multicolumn{1}{c}{}& $\Delta$                 & synaptic delay        & 10\,ms                                                                                     & 0\,ms (0\,ms)                                                                                            \\
\bottomrule
\end{tabular}
\end{center}
%* For all the populations $\alpha$ and for the two populations $\beta$ that are excitatory, $J^{\alpha\beta}=9.984$; for all the populations $\alpha$ and for the population $\beta$ that are inhibitory, $J^{\alpha\beta}=-19.968$.
* i.e. for all population $\alpha$, $J^{\alpha\beta} = 9.984$ mV if $\beta$ is an excitatory population and $J^{\alpha\beta} = -19.968$ mV if $\beta$ is the inhibitory population.
\end{table}

\begin{table}[]
\centering
\caption{Performance summary (ii) when fitting neuLVMs to the single-population example (Section~\ref{sec:3pop}, Figure~\ref{fig:1pop-exp}C) with $m$-cluster states. For each ground truth $J$, $10$ different datasets were generated and tested. ($6$ observed neurons.)}
\label{tab:1pop-result-fit}
\begin{tabular}{cccccccc}
\toprule
$J$                 & 59    & 60    & 61    & 62    & 63    & 64    & 65    \\ \midrule
$\widehat{J}$ (mean)              & 58.18 & 60.46 & 61.20 & 62.05 & 62.39 & 63.58 & 63.66 \\
$\widehat{J}$ (std)               & 0.50  & 0.71  & 0.93  & 0.46  & 1.11  & 1.59  & 1.85  \\ \midrule
Pearson $r$        & \multicolumn{7}{c}{0.81 ($p= 2.8\mathrm{e}{-17}$)}                 \\ \bottomrule
\end{tabular}
\end{table}

\begin{table}[]
\centering
\caption{Performance summary (i) when fitting neuLVMs to the single-population example (Section~\ref{sec:3pop}, Figure~\ref{fig:1pop-exp}F) with a transition from a metastable 4-cluster state to a 3-cluster state. For each ground truth $J$, $10$ different datasets were generated and tested. ($J=60.32$ mV.)}
\label{tab:1pop-result-comp}
\begin{tabular}{cccccc}
\toprule
\# observed neurons & 1     & 2     & 5     & 10    & 599   \\\midrule
$\widehat{J}$ (mean)              & 60.37 & 60.14 & 59.33 & 59.81 & 59.10 \\
$\widehat{J}$ (std)               & 2.43  & 1.48  & 0.91  & 1.25  & 1.26  \\ \bottomrule
\end{tabular}
\end{table}

\begin{table}[]
% \centering
\caption{Hyper-parameters used when fitting neuLVM. }
\label{tab:hyper-fitting}
\begin{center}
\begin{tabular}{lll}
\toprule
Name             & \multicolumn{2}{c}{Value}                                                                                                                                                                       \\ \cmidrule(r){2-3} 
\multicolumn{1}{c}{}                          & \multicolumn{1}{c}{\begin{tabular}[c]{@{}c@{}}Example Section~\ref{sec:single}\\ Single excit. population\end{tabular}} & \multicolumn{1}{c}{\begin{tabular}[c]{@{}c@{}}Example Section~\ref{sec:3pop}\\ Excitat. (inhib.) populations\end{tabular}} \\ \midrule
$\mathrm{lr}_{\mathrm{E}}$                        & $1\mathrm{e}{-3}$                                                                                       & $1\mathrm{e}{-3}$                                                                                                 \\
$\mathrm{maxiter}_{\mathrm{E}}$                                    & 200                                                                                      & 200                                                                                            \\
% $\mathrm{ftol}_{\mathrm{E}}$                & $3\mathrm{e}{-4}$                                                                             & $3\mathrm{e}{-4}$                                                                                   \\
$\mathrm{itertol}_{\mathrm{E}}$           & 3                                                                                      & 3                                                                                            \\
$\mathrm{lr}_{\mathrm{M}}$                        & $1\mathrm{e}{-8}$ *                                                                                      & $1\mathrm{e}{-8}$*                                                                                                 \\
$\mathrm{maxiter}_{\mathrm{M}}$                 & 200                                                                                     & 200                                                                                         \\
$\mathrm{ftol}_{\mathrm{M}}$                               & $2\mathrm{e}{-9}$*                                                                                      & $2\mathrm{e}{-9}$*                                                                                            \\
$\mathrm{gtol}_{\mathrm{M}}$                           & $1\mathrm{e}{-5}$*                                                                          & $1\mathrm{e}{-5}$*                                                                               \\
\bottomrule
\end{tabular}
\end{center}
* Values are default as in scipy.optimize.minimize(method=`L-BFGS-B').
\end{table}

% \begin{figure}
% \centering
% \includegraphics[height=\textheight]{Figures/Appendix_transition.png}
% \caption{}\label{fig:ap-transition}
% \end{figure}

\begin{table}[]
\centering
\caption{Within-population heterogeneity introduced in the ground truth winner-take-all (WTA) network (in the additional experiment of Appendix~\ref{app:additional_exp}).}
\label{tab:3pop-result-hetero}
\begin{tabular}{ccc}
\toprule
ground truth within-population heterogeneity & $\mu$     & $\sigma$ (normal distribution)     \\\midrule
$J^{\textrm{e}^1} J^{\textrm{e}^2}, J^{\textrm{i}}$              & 9.98 / 9.98 / 19.97	 & 2.00 / 2.00 / 2.00 \\
$\vartheta$                & 3.70  & 0.07 \\ 
$U_\mathrm{r}$                & 14.40 & 0.29 \\ 
$\tau_{\mathrm{mem}}$                & 20.00  & 0.40 \\ \bottomrule
\end{tabular}
\end{table}

\newpage

\section{Checklist}
\begin{enumerate}

\item For all authors...
\begin{enumerate}
  \item Do the main claims made in the abstract and introduction accurately reflect the paper's contributions and scope?
    \answerYes{}
  \item Did you describe the limitations of your work?
    \answerYes{}
  \item Did you discuss any potential negative societal impacts of your work?
     \answerNA{}
  \item Have you read the ethics review guidelines and ensured that your paper conforms to them?
    \answerYes{}
\end{enumerate}

\item If you are including theoretical results...
\begin{enumerate}
  \item Did you state the full set of assumptions of all theoretical results?
    \answerYes{}
        \item Did you include complete proofs of all theoretical results?
    \answerYes{}
\end{enumerate}

\item If you ran experiments...
\begin{enumerate}
  \item Did you include the code, data, and instructions needed to reproduce the main experimental results (either in the supplemental material or as a URL)?
    \answerYes{} Code URL available upon acceptance or upon request.
  \item Did you specify all the training details (e.g., data splits, hyperparameters, how they were chosen)?
    \answerYes{}
        \item Did you report error bars (e.g., with respect to the random seed after running experiments multiple times)?
    \answerYes{}
        \item Did you include the total amount of computing and the type of resources used (e.g., type of GPUs, internal cluster, or cloud provider)?
    \answerNA{}
\end{enumerate}

\item If you are using existing assets (e.g., code, data, models) or curating/releasing new assets...
\begin{enumerate}
  \item If your work uses existing assets, did you cite the creators?
    \answerYes{}
  \item Did you mention the license of the assets?
    \answerNA{}
  \item Did you include any new assets either in the supplemental material or as a URL?
    \answerNA{}
  \item Did you discuss whether and how consent was obtained from people whose data you're using/curating?
    \answerNA{}
  \item Did you discuss whether the data you are using/curating contains personally identifiable information or offensive content?
    \answerNA{}
\end{enumerate}

\item If you used crowdsourcing or conducted research with human subjects...
\begin{enumerate}
  \item Did you include the full text of instructions given to participants and screenshots, if applicable?
    \answerNA{}
  \item Did you describe any potential participant risks, with links to Institutional Review Board (IRB) approvals, if applicable?
    \answerNA{}
  \item Did you include the estimated hourly wage paid to participants and the total amount spent on participant compensation?
    \answerNA{}
\end{enumerate}
\end{enumerate}

\bibliographystyleSM{unsrtnat}
\bibliographySM{mybib}
\end{document}